\documentclass[prl,aps,amsfonts,preprint,floats,tightenlines,floatfix]{revtex4}
\usepackage{ifthen}                   
\usepackage{exscale}                  
\usepackage[intlimits]{amsmath}       
\usepackage{amsfonts}
\usepackage{amssymb,amscd}
\usepackage[dvips]{epsfig}                   
\usepackage{array}
\usepackage{wrapfig}

\newcommand{\cS}{\mathcal{S}}

\newcommand{\beqa}{\begin{eqnarray}}
\newcommand{\eeqa}{\end{eqnarray}}
\newcommand{\beq}{\begin{equation}}
\newcommand{\eeq}{\end{equation}}
 
\begin{document}
\date{\today}
\title{
\hspace*{\fill}{\small\sf UNITU--THEP--25/2002  FAU-TP3-02/27}\\
\hspace*{\fill}{\small\sf http://xxx.lanl.gov/abs/hep-th/0304134}\\ ~\\ ~\\
On the infrared behaviour of Gluons and Ghosts in\\
Ghost-Antighost symmetric gauges}

\author{
R.~Alkofer\footnote{E-Mail: reinhard.alkofer@uni-tuebingen.de}, 
C.~S.~Fischer\footnote{E-Mail: C.Fischer@thphys.uni-heidelberg.de\\
Address after April 1st, 2003:\\ Institute for Theoretical Physics,
Heidelberg University, Philosophenweg 16, D-69120 Heidelberg, Germany}, 
H.~Reinhardt\footnote{E-Mail: hugo.reinhardt@uni-tuebingen.de}}
\affiliation{Institute for Theoretical Physics, University of T\"ubingen \\
          Auf der Morgenstelle 14, D-72076 T\"ubingen, Germany}
\author{L.~von Smekal\footnote{E-Mail: smekal@theorie3.physik.uni-erlangen.de}}
\affiliation{Institute for Theoretical Physics III, 
             University of Erlangen-N\"urnberg \\
             Staudtstr.~7, D-91058 Erlangen, Germany}

\bigskip

\begin{abstract}
To investigate the possibility of a ghost-antighost condensate
the coupled Dyson--Schwinger equations for the gluon and ghost propagators in
Yang--Mills theories are derived in general covariant gauges,
including ghost-antighost symmetric gauges.  
The infrared behaviour of these two-point functions is studied in a bare-vertex
truncation scheme which has proven to be successful in Landau gauge. In all
linear covariant gauges the same infrared behaviour as in Landau gauge is
found: The gluon propagator is infrared suppressed whereas the ghost propagator
is infrared enhanced. This infrared singular behaviour provides indication
against a ghost-antighost condensate. 
In the ghost-antighost symmetric gauges we find that the infrared
behaviour of the gluon and ghost propagators cannot be determined when
replacing all dressed vertices by bare ones. The question of a BRST
invariant dimension two condensate remains to be further studied.

\bigskip

Gluon propagator; Dyson--Schwinger equations; Infrared behavior.

\bigskip


\end{abstract}
\maketitle

\section{Introduction}

A large body of experimental data supports the general believe that Quantum
Chromodynamics (QCD) is the correct theory of strong interactions. Nevertheless
we are left with the task of understanding the physics of hadrons, and hereby 
in particular the mechanisms of confinement and spontaneous breaking of chiral
symmetry. Gaining such an insight requires reliable non-perturbative treatments
of QCD. Hereby Monte Carlo lattice calculations provide a rigorous
non-perturbative approach to QCD. They have the advantage of fully respecting
gauge invariance independently of the size of the lattice used. On the other
hand, the extraction of the continuum values of physical observables from the
lattice data requires a careful study of the scaling regime. The observed
scaling behaviour, however, will be in general contaminated by finite size
effects. With respect to studies of the confinement mechanisms this is 
problematic: As infrared singularities are expected to occur in QCD there is
definite need for a continuum-based non-perturbative approach.

To this end we note that the Schwinger--Dyson equations of QCD can address
directly the infrared region. They provide genuine non-perturbative information
and are at the same time fully formulated in the continuum theory. Such an
approach is, however, less rigorous than lattice calculations in the sense that
truncations of the tower of coupled equations are necessary in practical
calculations. Justifications for such truncations can be given on the
basis of general principles like {\it e.g.\/} a restriction to the first Gribov
region, see ref.\ \cite{Zwanziger:2003cf} and references therein. Nevertheless,
the validity of the employed truncation is finally judged by comparing its
results with either the results of Monte Carlo calculations or experiments. The
latter is easily possible as the Schwinger--Dyson approach has been
successfully applied to the description of hadron phenomenology, see {\it e.g.}
the recent reviews  refs.\ \cite{Roberts:2000aa,Alkofer:2001wg} and references
therein. Furthermore, despite recent progress by improved lattice algorithms,
and despite the increasing computer time available for lattice calculations,
including dynamical fermions is exceedingly cumbersome and finite baryon
densities are hardly accessibly in realistic SU(3) lattice simulations.  On the
other hand, dynamical fermions and finite baryon densities can be relatively
easily treated in the Schwinger--Dyson approach to QCD. 

In recent years  the fundamental Schwinger--Dyson equations of SU(N) Yang-Mills
theories have been solved explicitly  in certain approximations yielding gluon
and ghost propagators 
\cite{Alkofer:2001wg,vonSmekal:1997is,Atkinson:1998tu,Zwanziger:2001kw,
Lerche:2002ep,Fischer:2002hn}. In these calculations, carried
out in Landau gauge, vertex functions constructed from appropriate
Slavnov-Taylor identities as well as bare vertices have been employed. The
results proved to be qualitatively similar among each other and agree well with
recent lattice calculations 
\cite{Bonnet:2000kw,Bonnet:2001uh,Langfeld:2001cz,Suman:1995zg,
Cucchieri:1997dx} for both, the gluon and ghost propagator.  The common, though
gauge dependent, result of both approaches is an infrared suppressed gluon
propagator and an infrared enhanced ghost propagator. Furthermore, the
inclusion of dynamical quarks does not alter the infrared behaviour of gluon
and ghost propagators and leads to only slight modifications for
non-vanishing 
momenta for the number of light flavours $N_f \le 3$  \cite{Fischer:2003rp}.
These results especially imply that the ghosts take the role of the long range
correlations in the theory. Such a behaviour is in accordance with the
Gribov--Zwanziger horizon condition, see {\it e.g.} ref.\
\cite{Zwanziger:2001kw} and references therein, and the Kugo--Ojima confinement
criterion, which in Landau gauge includes the statement that the ghost
propagator should be more singular than a simple pole \cite{Kugo:1979gm}. 

The central assumption in the Kugo--Ojima confinement scenario is the
invariance of the measure of  the functional integral under BRS transformations
and the existence of a nilpotent  BRS-operator \cite{Nakanishi}. The most
general Lorentz invariant and globally gauge invariant Lagrangean of dimension
four that can be constructed under this assumption has been derived in ref.\
\cite{Baulieu:1982sb}. In addition to the structure appearing in ordinary linear
covariant gauges, the Lagrangean contains a second gauge parameter which
controls the symmetry of the Lagrangean under ghost-antighost interchange. 
Furthermore a four ghost interaction term is present. 
We will use this Lagrangean as the starting point of our investigation. 

Our main interest in this paper will be to explore the situation in
these general covariant gauges. Away from the Landau gauge limit the
connection between the Kugo--Ojima confinement criterion and the
infrared behaviour of the ghost dressing function is far from obvious. 
In particular, the question might arise whether it is possible that
the infrared dominant role of the ghost dressing function, seen in the
Landau gauge, is assumed by other degrees of freedom like the
longitudinal gluons in other covariant gauges. As a matter of fact, infrared
dominance of  longitudinal gluons is seen if stochastic quantization is used
instead of the  Faddeev--Popov quantization \cite{Zwanziger:2002ia}.
Furthermore,  calculations based on many-body techniques provide evidence  that
in Coulomb gauge (employing the usual Faddeev--Popov quantization) the ghosts
and the Coulomb gluons are both infrared enhanced \cite{Szczepaniak:2001rg}.
This latter  picture for Coulomb gauge QCD obtains (at least partial) support
from lattice \cite{Cucchieri:2001zb} and renormalization group calculations
\cite{Cucchieri:2000hv}. Care has, however, to be taken as the Coulomb gauge
limit is highly non-trivial, see ref.\ {\it e.g.} \cite{Baulieu:1998kx}. On the
other hand, the benefit of Coulomb gauge is obvious. The time-time component of
the gluon propagator and the heavy quark potential fulfill a strictly valid
inequality \cite{Cucchieri:2000hv,Zwanziger:2002sh} with the Coulomb string
tension being several times larger than the asymptotic one
\cite{Greensite:2003xf}. Even more important, quark confinement directly
results from infrared enhanced Coulomb gluons, see {\it e.g.} refs.\
\cite{Szczepaniak:1996tk,Alkofer:tc} and references therein. Instead of
exploring the correlation functions in non-covariant gauges we will in this
paper study Green's functions in covariant albeit non-linear gauges.

Ghost-antighost symmetric gauges are of special interest when investigating 
the possibility of a BRST invariant condensate of dimension two in QCD. Such
condensates occur in the operator product expansion of the gluon propagator
\cite{Lavelle:eg,Boucaud:2001st,Kondo:2001nq}, bear some relation to the Gribov problem
\cite{Stodolsky:2002st}, may result in gluon mass generation
\cite{Dudal:2003gu} and may be important for confinement in general
\cite{Schaden:1999ew,Gubarev:2000eu}. Hereby it has been clarified recently 
that these condensates are highly non-local \cite{Dan, Pierre} and that they
are only BRST invariant after eliminating the Nakanishi--Lautrup field via its
equation of motion \cite{Gripaios:2003xq}. This kind of restricted BRST
invariance has been called `on-shell BRST invariance' and can be related to
a residual gauge symmetry after gauge fixing. 

The solutions of the gluon and ghost Dyson--Schwinger equations in Landau gauge
provide a somewhat different picture: Whereas the operator product expansion of
the gluon propagator requires such a dimension two condensate its
interpretation with respect to a gluon mass is made impossible by the gluon
propagator's infrared behaviour $D(p^2=0)=0$ instead of $D(p^2=0)=1/m^2$. Also
the highly infrared singular ghost propagator excludes a ghost mass and/or a
ghost-antighost condensate. Therefore the question arises whether in general 
ghost-antighost symmetric gauges the infrared behaviour of the propagators 
can be interpreted in terms of gluon and ghost ``masses''.

This paper is organized as follows: In sect.~2 we summarize some properties of
the general Lagrangean given in ref.\ \cite{Baulieu:1982sb} and outline the
derivation of the coupled set of Dyson--Schwinger equations (DSEs) for the
ghost and gluon propagators. As the Lagrangean contains a four--ghost
interaction a rich structure in the ghost DSE emerges which closely resembles
the one already present in the gluon equation of ordinary linear covariant
gauges. In sect.~3 we employ a truncation scheme that has proven to be
successful in Landau gauge and study in particular the infrared behaviour of
the ghost and gluon dressing functions for general values of the two gauge
parameters.  Furthermore, we show that in the ghost-antighost symmetric
gauges the contributions of  the genuine two-loop terms (generalized
squint and sunset diagram) in the gluon and the ghost DSEs must be
properly taken into account in the infrared. In the linear covariant
gauges no such terms are present in the ghost DSE, 
and selfconsistent results can be obtained assuming the two-loop terms in the gluon
equation to be subleading in the infrared \footnote{This has been demonstrated 
in refs.~\cite{Zwanziger:2001kw,Lerche:2002ep,Fischer:2002hn}. 
On the other hand selfconsistent solutions can also be obtained
once the two-loop diagramms are assumed to contribute in the infrared \cite{Bloch:2003}.}.  
In general 
ghost-antighost symmetric gauges, on the other hand, the bare-vertex truncation
is insufficient to clarify the infrared behaviour of the gluon and ghost
propagators. In sect.~4 we will provide numerical solutions for the DSEs in the
Landau gauge limit of the ghost--antighost symmetric case of the Lagrangean and
recover the solutions found in \cite{Fischer:2002hn} from a different direction
in two dimensional gauge parameter space. In the last section we give our
conclusions. Technical details are deferred into four appendices.

\section{The Dyson--Schwinger equation for the ghost propagator}

\subsection{Renormalized double BRS symmetry}

The most general Lagrangean of dimension four that is  Lorentz invariant,
globally gauge invariant, invariant under BRST- and anti-BRST-transformations,
hermitean and omitting topological terms, is \cite{Baulieu:1982sb}:
\beqa
\cal{L} &=& \frac{1}{4} F_{\mu \nu}^2 + \frac{\left(\partial_\mu A_\mu \right)^2}
{2 \lambda} 
+ \frac{\alpha}{2} \left(1-\frac{\alpha}{2}\right) \frac{\lambda}{2}
(\bar{c}\times c)^2 
- i\frac{\alpha}{2} D_\mu \bar{c} \partial_\mu c
- i\left(1-\frac{\alpha}{2} \right) \partial_\mu \bar{c} D_\mu c .
\label{Lagrangian}
\eeqa
The field strength tensor and the covariant derivative are defined as
\beqa
F_{\mu \nu}^a &=& \partial_\mu A_\nu^a - \partial_\nu A_\mu^a -g f^{abc} A_\mu^b
A_\nu^c \nonumber\\
D_\mu^{ab} &=& \partial_\mu \delta^{ab} + g f^{abc} 
A_\mu^c ,
\eeqa
and the abbreviation $(\bar{c} \times c)^a= gf^{abc} \: \bar{c}^b c^c$ is used.
Note that both ghost and antighost fields, $\bar{c}$ and $c$, resp., are chosen
to be hermitean, $c^\dagger = c$ and $\bar c^\dagger = \bar c$ . This is
necessary to maintain the hermiticity of the Lagrangian for all values of the
gauge parameters $\lambda$ and $\alpha$, see {\it e.g.} \cite{Nakanishi} and
references therein. Furthermore we work in Euclidean space-time. 

From the two gauge parameters of the Lagrangian the first one, $\lambda$, is
the  usual parameter of linear covariant gauges, whereas the second one,
$\alpha$, controls the symmetry properties of the ghost content. For the cases
$\alpha=0$ and $\alpha=2$ one recovers the usual Faddeev--Popov Lagrangian and
its mirror image, respectively, where the role of ghost and antighost have been
interchanged. For the value $\alpha=1$ the Lagrangian is completely symmetric
in the ghost and antighost fields.

In ref.\ \cite{Baulieu:1982sb} it has been shown that the S-matrix of the
theory is invariant under variation of the gauge parameters $\lambda$ and
$\alpha$. Therefore gauge invariance of physical observables is ensured.
One-loop calculations confirm in particular the independence of the first
nontrivial coefficient of the $\beta$ function from the gauge parameters.

Furthermore, the existence of a renormalized BRS-algebra has been proven
\cite{Baulieu:1982sb}, thus the theory given by (\ref{Lagrangian}) is
multiplicatively renormalizable. From one-loop calculations one finds, that the
Faddeev-Popov values of the gauge parameters,  $\alpha=0$ and $\alpha=2$, are
fixed points under the renormalization procedure. The same is true for the
ghost-antighost symmetric case $\alpha=1$. The case of Landau gauge,
$\lambda=0$,  corresponds to a fixed point as well, because the constraint
$\partial_\mu A^\mu=0$ is not affected  by a rescaling of the gluon field.  

To be specific the renormalized BRS ($s_r$) and anti-BRS
($\bar s_r$) transformations are given by
\begin{equation}
\begin{array}{l@{\hskip 20mm}l}
s_r A = - \widetilde Z_3 D_r c \; , &  
\bar s_r A = - \widetilde Z_3 D_r \bar c \; ,\\
&\\
\displaystyle 
s_r c = - \widetilde Z_1 \frac{1}{2} \, (c \times c ) \; ,&
\displaystyle 
\bar s_r \bar c = - \widetilde Z_1  \frac{1}{2} \, (\bar c\times \bar c) \; , \\
&\\
\displaystyle 
s_r \bar c = B - \frac \alpha 2 \widetilde Z_1   \, (\bar c\times c) \; ,  & 
\displaystyle
\bar s_r c = - B - (1-\frac \alpha 2 ) \widetilde Z_1   \, (\bar c\times c) 
\; , \\
&\\
\displaystyle  
s_r B =  - \frac \alpha 2  \widetilde Z_1   \, (c\times B ) &
\displaystyle
\bar s_r B = - (1-\frac \alpha 2 ) \widetilde Z_1   \, (\bar c\times B ) \\
\displaystyle 
\qquad - \frac \alpha 2 (1-\frac \alpha 2 )
\frac{1}{2} {\widetilde Z}_1^2  \, \Big( ( c \times c ) \times\bar c\Big) \; ,&
\displaystyle  
\qquad  +\frac \alpha 2 (1-\frac \alpha 2 )
\frac{1}{2} {\widetilde Z}_1^2  \, \Big( (\bar c \times \bar c )  
\times c\Big) \; .
\end{array} 
\label{realBRS}
\end{equation}
Here $D_r = \partial - Z_3^{1/2} Z_g  ( A \times \phantom{A}) $
is the covariant derivative in the adjoint representation,
with color and Lorentz indices suppressed. 
Note that the Nakanishi--Lautrup auxiliary field $B$ can be eliminated from the
BRS-transformations by using its equation of motion.
The corresponding BRS-transformations are called `on-shell'.
Note furthermore that the application of the BRS-operator $s_r$ ($\bar s_r$)
on a field increases (decreases)
the ghost number by $+1$ ($-1$), thus we can assign the 
value $N_{FP}=+1$ ($N_{FP}=-1$) to the (anti-)BRS-operator itself.
The BRS-operator and the anti-BRS-operator are nilpotent and related by
$s_r\bar{s}_r + \bar{s}_rs_r = 0$. These properties are, however, lost when
considering `on-shell' BRS-transformations.

The Maurer-Cartan conditions, in addition to the forms of $s_r c$ and 
$\bar s_r \bar c$, for ghosts $c$ and anti-ghosts $\bar c$  in a ghost 
anti-ghost symmetric formulation thereby require \cite{Baulieu:1982sb} 
\begin{equation}
 s_r \bar c +  \bar s_r c  + \widetilde Z_1  \, (\bar c\times c) = 0 \; .
\end{equation}

The correspondence between the bare Lagrangean and its renormalized version
including counterterms is given by the following rescaling transformations
\beqa
A_\mu^a &\rightarrow& \sqrt{Z_3}A_\mu^a, 
\:\:\:\:\:\: \bar{c}^a c^b \rightarrow \tilde{Z}_3 \bar{c}^a c^b,
\:\:\:\:\:\:  B^a \rightarrow B^a/\sqrt{Z_3} \nonumber \\ 
g &\rightarrow &Z_g g, \:\:\:\:\:\: \:\:\:\:\:\: \:
\alpha \rightarrow Z_\alpha \alpha, \:\:\:\:\:\: \:\:\:\:\:\:\:\:
\lambda \rightarrow Z_\lambda \lambda, 
\label{renorm}
\eeqa
where five independent renormalization constants $Z_3,\tilde{Z}_3,Z_g,Z_\alpha$
and $Z_\lambda$ have been introduced. Furthermore four additional
renormalization constants are related to these via  Slavnov--Taylor identities,
\beq
Z_1 = Z_g Z_3^{3/2}, \:\:\:\:\:\: \tilde{Z}_1=Z_g \tilde{Z}_3 Z_3^{1/2}, 
\:\:\:\:\:\: Z_4=Z_g^2 Z_3^2,
\:\:\:\:\:\: \tilde{Z}_4=Z_g^2 \tilde{Z}_3^2 .
\eeq
Note, however, that contrary to standard Faddeev--Popov gauges 
$Z_\lambda \not=  Z_3$, {\it e.g.}, at one-loop (MS scheme), one has 
\footnote{Note the typos in \protect\cite{Baulieu:1982sb}. 
For explicit calculations of the renormalization
constants for the ghost anti-ghost symmetric case $\alpha=1$
with Curci-Ferrari mass term and quarks, up to three loops, see
\cite{deBoer:1995dh,Browne:2002wd}.}
\begin{equation}
  \label{Zxi}
  Z_\lambda  = Z_3 - \frac{g^2}{16\pi^2}  \frac{1}{\epsilon} N_c \,
              \frac \alpha 2 (1-\frac \alpha 2 ) \; \lambda    \; .
\end{equation}

The gauge fixing part of the Lagrangean (\ref{Lagrangian}) can be written 
in the following three equivalent ways,
\begin{eqnarray}
       \mathcal{L}_{\mathcal{GF}} &=& \frac{i}{2} \frac{1}{Z_3\widetilde Z_3}
\, s_r \bar s_r \left(   Z_3\,  AA + i Z_\lambda  \widetilde Z_3 \,  \alpha \,
\lambda  c \bar c \right) \, + \, \frac{Z_\lambda }{Z_3} (1-\alpha)  
\frac{\lambda}{2}
         \, s_r (\bar c \, s_r \bar c ) \label{Lgf1} \\
&=& i s_r \left( \bar  c \big( \partial A - i \frac{Z_\lambda}{Z_3} 
\frac{\lambda}{2}
B \big) \right)  \label{Lgf2} \\
&=& i B\partial A +   \frac{Z_\lambda}{Z_3} \frac{\lambda}{2} B^2 +
\frac{Z_\lambda}{Z_3} {\widetilde Z}_1^2 \,  
\frac \alpha 2 (1-\frac \alpha 2 ) \,
\frac{\lambda}{2}  (\bar c \times c )^2 + i \widetilde Z_3 
\big((1-\frac \alpha 2 ) \, \bar c \, \partial D_r \, c \, +
\, \frac \alpha 2 \, \bar c\, D_r\partial \,  c \big) \; . 
\nonumber \\  \label{Lgf3}
\end{eqnarray}
This is verified by direct calculation via the transformations defined in
Eqs.~(\ref{realBRS}). In the form of Eq.~(\ref{Lgf3}) the gauge fixing
Lagrangean shows that the renormalization constants introduced in
(\ref{realBRS}) correspond to the replacements of bare by renormalized
quantities as given above.

We may rewrite the gauge fixing Lagrangean of Eq.~(\ref{Lgf3}) once more,
\begin{eqnarray}
  \label{Lgf4}
     \mathcal{L}_{\mathcal{GF}} &=&
 i B\partial A +   \frac{Z_\lambda}{Z_3} \frac{\lambda}{2} B^2 +
\frac{Z_\lambda}{Z_3} {\widetilde Z}_1^2 \, 
\frac \alpha 2 (1-\frac \alpha 2 ) \,
 \frac{\lambda}{2} (\bar c \times c )^2 + i \widetilde Z_3  \frac{1}{2}
\big( \bar c \, \partial D_r \, c \, +\, \bar c\, D_r\partial \,  c \big)
\nonumber \\
&+& i \widetilde Z_1 (1- \alpha )\, \frac{1}{2}  \, \partial A \,
 ( \bar c \times c)
 \; .
\end{eqnarray}
This emphasizes the role of the gauge parameter $\alpha$. In this form,  the
only term not symmetric under Faddeev--Popov conjugation, $c \to \bar c$ and $
\bar c \to - c$, is the last one (which is anti-symmetric w.r.t.\ 
Faddeev--Popov conjugation). It vanishes for $\alpha\!=\!1$. With the current
(real) hermiticity assignment for ghost and anti-ghost fields the Lagrangean
is hermitean for all $\alpha$, and it reduces to the standard  Faddeev--Popov
form for $\alpha=0$.  We could also introduce hermitean adjoint ghost and
anti-ghost fields, with the assignment $c^\dagger = \bar c$, via the Caley map
\footnote{See the appendix of Ref.~\cite{Alkofer:2001wg}}.  This would then
lead to
\begin{eqnarray}
  \label{Lgf5}
     \mathcal{L}^{cc}_{\mathcal{GF}} &=&
 i B\partial A +   \frac{Z_\lambda}{Z_3} \frac{\lambda}{2} B^2 -
\frac{Z_\lambda}{Z_3} {\widetilde Z}_1^2 \,  
\frac \alpha 2 (1-\frac \alpha 2 )  \,
 \frac{\lambda}{2} (\bar c \times c )^2 +  \widetilde Z_3  \frac{1}{2}
\big( \bar c \, \partial D_r \, c \, +\, \bar c\, D_r\partial \,  c \big)
\nonumber \\
&+& \widetilde Z_1 (1-\alpha )\, \frac{1}{4} \,  \big( \bar c
 \,  (\partial A  \times \bar c) -   c\,  (\partial A  \times c)
 \big)     \; .
\end{eqnarray}

While this form of the Lagrangean, which we will not use further herein,  is
still hermitean it no longer reduces to the form of standard
Faddeev--Popov theory for $\alpha=0$. Thus the Faddeev--Popov Lagrangean
is only consistent with hermiticity for the choice of real ghost fields
\cite{Nakanishi}.
With complex conjugate ghost and anti-ghost fields, additional terms
for $\alpha =0$ survive (which are absent in standard Faddeev--Popov gauges). 
Only for $\alpha \! =\! 1$ both versions, with hermitean real or
complex conjugate ghost pairs, have the same Lagrangean and may be
interchanged arbitrarily.

\subsection{Ghost and Anti-Ghost Dyson--Schwinger equations}

Without invariance under Faddeev--Popov conjugation, {\it i.e.}~without
ghost anti-ghost symmetry ($\alpha =1$ or $\lambda=0$), we have separate  
ghost and anti-ghost DSEs which are {\bf not} identical.  Consider the following
representations of the ghost (anti-ghost) derivatives of the action (for
brevity we indicate by subscripts the space-time arguments of fields),
\begin{eqnarray}
  \label{ghDer}
  \frac{\delta S}{\delta c^a_x} &=& \widetilde Z_3 \, i (\partial D_r\bar
  c)^a_x \,+\,\frac{Z_\lambda \lambda}{Z_3}  \widetilde Z_1 \, 
  (1-\frac \alpha 2) \, (\bar c
  \times B)^a_x \, - \,  \frac{Z_\lambda \lambda}{Z_3} \,  
  {\widetilde Z}^2_1  \frac \alpha 4 (1-\frac \alpha 2 )
  \big((\bar c \times\bar c)\times c\big)^a_x 
\nonumber  \\
                         &=& \widetilde Z_3 \, i (\partial D_r\bar
  c)^a_x \, - \,\frac{Z_\lambda \lambda}{Z_3} \, 
  \bar s_r B^a_x \,=\,  -i \, \bar s_r
  \big( \partial A^a_x \, - \, i\frac{Z_\lambda \lambda}{Z_3} \, B^a_x \big) \; 
,\\
   \label{aghDer}
  \frac{\delta S}{\delta \bar c^a_x} &=& \widetilde Z_3 \, i (\partial D_r
   c)^a_x \,+\,\frac{Z_\lambda \lambda}{Z_3}  \widetilde Z_1   \, 
   \frac \alpha 2  \, (c
  \times B)^a_x \, + \,  \frac{Z_\lambda \lambda}{Z_3} \,  
  {\widetilde Z}^2_1   \frac \alpha 4 (1-\frac \alpha 2 )
 \big((c \times c)\times \bar c\big)^a_x 
\nonumber \\
                         &=& \widetilde Z_3 \, i (\partial D_r c)^a_x \, -
   \,\frac{Z_\lambda \lambda}{Z_3} \,  s_r B^a_x \,=\,  -i \, s_r
  \big( \partial A^a_x \, - \, i\frac{Z_\lambda \lambda}{Z_3} \, 
  B^a_x \big) \; .
\end{eqnarray}

The two DSEs then follow readily from
\begin{equation}
  \label{ghDSEs}
  \langle \, \frac{\delta S}{\delta\bar c^a_x} \, \bar c^b_y\, \rangle \,
    =\, \langle \, c^b_y \,  \frac{\delta S}{\delta c^a_x} \, \rangle
    \,=\, \delta^{ab} \, \delta_{xy} \; .
\end{equation}
Of course, they are related by Faddeev--Popov conjugation 
$\mathcal{C}_{\mbox{\tiny FP}}$
which interchanges the two. In particular,
\begin{equation}
  \label{eq:CFP}
    \mathcal{C}_{\mbox{\tiny FP}} c = \bar c \; , \;\;
    \mathcal{C}_{\mbox{\tiny FP}} \bar c =  - c \; , \;\;
    \mathcal{C}_{\mbox{\tiny FP}} B = B + \widetilde Z_1  (1-\alpha )
    (\bar c \times c)  \; , \;\;  \mathcal{C}_{\mbox{\tiny FP}} A = A \;.
\end{equation}
The transformation of the Nakanishi--Lautrup $B$-field follows from
compatibility with BRS/anti-BRS invariance and,
\begin{equation}
  \label{sbs}
  \bar s_r \,=  \,  \mathcal{C}_{\mbox{\tiny FP}}  \, s\,
  \mathcal{C}_{\mbox{\tiny FP}}^{-1}  \; .
\end{equation}
On the level of the BRS and anti-BRS transformations we can have this form of
Faddeev--Popov conjugation for arbitrary $\alpha$. However, it is relatively
easy to verify that the Lagrangean, {\it i.e.\/} the measure of the theory,
is not invariant under $\mathcal{C}_{\mbox{\tiny FP}}$ and thus ghost and
antighost DSEs are not identical, unless $\alpha = 1$ or $\lambda=0$:
With the above Faddeev--Popov conjugation rule for the $B$-field, the sign 
change in the
last term of (\ref{Lgf4}) is exactly compensated by the first term,
\begin{eqnarray}
  \label{compens}
 i B\partial A + i \widetilde Z_1 (1-\alpha )\, \frac{1}{2}  \,
 \partial A \,  ( \bar c \times c)
  &\stackrel{\mathcal{C}_{\mbox{\tiny FP}}}{\longrightarrow}&
 i\big( B + \widetilde Z_1  (1-\alpha )
    (\bar c \times c)\big) \partial A - i \widetilde Z_1 (1-\alpha)\,
 \frac{1}{2}  \, \partial A \,  ( \bar c \times c) \nonumber \\
 &=& i B\partial A + i \widetilde Z_1 (1-\alpha)\, \frac{1}{2}  \,
 \partial A \,  ( \bar c \times c) \;.
\end{eqnarray}
In this way, the violations of Faddeev--Popov conjugation invariance can
entirely be moved into the term $\propto \lambda B^2 $, and they thus obviously
disappear in the Landau gauge $\lambda\!=\!0$. On the other hand, in the more
general ghost anti-ghost symmetric case, with $\alpha =1$ and
$\mathcal{C}_{\mbox{\tiny FP}} B = B $, the theory does have the invariance
under Faddeev--Popov conjugation for all $\lambda$ and we can then immediately
conclude that expectation values of $ \mathcal{C}_{\mbox{\tiny FP}}$-odd
operators vanish.

Let us now look at one of the ghost DSEs, {\it e.g.}, from Eq.~(\ref{aghDer})
we obtain
\begin{equation}
  \label{ghDSE1}
   \delta^{ab} \, \delta_{xy}
\,   =\, \langle \, \frac{\delta S}{\delta\bar c^a_x} \, \bar c^b_y\, \rangle
\,   =\,  \widetilde Z_3 \,  \langle  \, i (\partial D_r c)^a_x \, \bar c^b_y\,
  \rangle \,  - \,\frac{Z_\lambda \lambda }{Z_3} \,\langle  \, (s_r B^a_x ) \, \bar
  c^b_y\, \rangle   \; .
\end{equation}
For the second term on the r.h.s.\ we write,
\begin{equation}
  \label{auxDSE1}
        \langle  \, (s_r B^a_x ) \, \bar
  c^b_y\, \rangle   \, =\,   \langle  \, s_r (B^a_x  \, \bar
  c^b_y ) \, \rangle   \, - \,    \langle  \, B^a_x  \,(s_r  \bar
  c^b_y ) \, \rangle   \, = \, - \langle \,  B^a_x \,  B^b_y \, \rangle \, + \,
  \widetilde Z_1\, (1-\frac \alpha 2 )  \,  \langle \,  B^a_x \,  
  (\bar c \times c )^b_y
  \, \rangle  \; ,
\end{equation}
where we have used that expectation values of total BRS variations vanish.
For the $B$-field correlations, and with its equation of motion
$ Z_\lambda \lambda \, B \,=\, i Z_3 \partial A$, one furthermore has,
\begin{equation}
  \label{BB}
  \frac{Z_\lambda \lambda}{Z_3} \,  \langle \,  B^a_x \,  B^b_y \, \rangle 
  \, = \,
  \delta^{ab} \, \delta_{xy} \, - \,  \frac{Z_3}{Z_\lambda \lambda} \,  \langle \,
      \partial A^a_x \, \partial A^b_y \, \rangle \; .
\end{equation}
Inserting Eqs.~(\ref{auxDSE1}) and (\ref{BB}) into the ghost DSE
(\ref{ghDSE1}) we arrive at
\begin{equation}
  \label{ghDSE2}
   \frac{Z_3}{Z_\lambda \lambda} \,  \langle \,
      \partial A^a_x \, \partial A^b_y \, \rangle \, = \,  
       \widetilde Z_3 \,  \langle  \, i (\partial D_r c)^a_x \, \bar c^b_y\,
  \rangle \,  + \, i\widetilde Z_1 \, (1-\frac \alpha 2) \, 
  \langle \, \partial A^a_x \,
      (\bar c \times c )^b_y \, \rangle \;.
\end{equation}
In the last term herein we inserted the e.o.m.\ for the $B$-field again. This
term is odd under Faddeev--Popov conjugation and thus vanishes in the 
ghost/anti-ghost symmetric case $\alpha =1$, as asserted above.
We thus have the important form of the ghost DSE in the Faddeev--Popov 
symmetric formulation (in which there is only one such DSE),
\begin{equation}
  \label{ghDSEs1}
   \frac{Z_3}{Z_\lambda \lambda} \,  \langle \,
      \partial A^a_x \, \partial A^b_y \, \rangle \, = \,   
      \widetilde Z_3 \,  \langle  \, i (\partial D_r c)^a_x \, \bar c^b_y\,
  \rangle  \;.
\end{equation}
Note that we obtain the same equation for standard Faddeev--Popov theory
($\alpha = 0$). The important difference to the standard form of the ghost DSE
is given by
\begin{equation}
Z_3 \,  \langle \,
    \partial A^a_x \, \partial A^b_y \, \rangle \, - \, {Z_\lambda \lambda} \,
      \delta^{ab} \, \delta_{xy} 
\end{equation}       
which vanishes in the usual Faddeev--Popov theory.
For general $\alpha $, however, the Slavnov--Taylor identities are modified 
also and this contribution does no longer need to vanish as we will see at
the end of this section.
Before that, we give a convenient (symmetrised) form of the
ghost DSE valid for arbitrary $\alpha$ without ghost anti-ghost invariance.
Note that we could equally have started from the ghost derivative in
Eq.~(\ref{ghDer}) and $ \langle \, c^b_y \,  (\delta /{\delta c^a_x} ) S \,
\rangle  \,=\, \delta^{ab} \, \delta_{xy} $. This would lead us to the 
Faddeev--Popov conjugate of Eq.~(\ref{ghDSE2}) (obtained from (\ref{ghDSE2})
with $c\to\bar c$, $\bar c \to -c$ and $\alpha \to 2-\alpha $).
Adding the two, we obtain a Faddeev--Popov symmetric
version in the place of Eq.~(\ref{ghDSE2}),
\begin{equation}
  \label{ghDSEs2}
   \frac{Z_3}{Z_\lambda \lambda} \,  \langle \,
      \partial A^a_x \, \partial A^b_y \, \rangle \, = \,   \widetilde Z_3 \,
       \frac{1}{2} \, \Big( \langle  \, i (\partial D_r c)^a_x \, \bar c^b_y\,
  \rangle \,  + \,  \langle  \,  c^b_y\,  i(\partial D_r \bar c)^a_x \,
  \rangle \Big) \, - \,
 i\widetilde Z_1 \,\frac{1}{2} \, (1-\alpha )  \, \langle \, \partial
      A^a_x \,       (\bar c \times c )^b_y \, \rangle \;.
\end{equation}
Just as we have a doubling of ghost DSEs, in absence of Faddeev--Popov 
conjugation invariance, we also have a doubling of Slavnov--Taylor identities.
As the result of one such new Slavnov--Taylor identity we will derive  
below that
\begin{equation}
  \label{auxSTI}
  \widetilde Z_1 \,\frac{1}{2} \, \, \langle \, \partial
      A^a_x \, (\bar c \times c )^b_y \, \rangle  \, = \,
   \widetilde Z_3 \,
       \frac{i}{2} \, \Big( \langle  \, i (\partial D_r c)^a_x \, \bar c^b_y\,
  \rangle \,  - \,  \langle  \,  c^b_y\,  i(\partial D_r \bar c)^a_x \,
  \rangle \Big) \;.
\end{equation}
This allows us to write for the ghost DSE (\ref{ghDSEs2}) and general $\alpha$,
finally,
\begin{equation}
  \label{ghDSEs3}
   \frac{Z_3}{Z_\lambda \lambda} \,  \langle \,
      \partial A^a_x \, \partial A^b_y \, \rangle \, = \,   \widetilde Z_3 \,
        \Big( (1-\frac \alpha 2) \, \langle  \, i (\partial D_r c)^a_x \, 
	\bar c^b_y\,
  \rangle \,  + \, \frac \alpha 2 \,  \langle  \,  c^b_y\,  
  i(\partial D_r \bar c)^a_x \,
  \rangle \Big)  \;.
\end{equation}
For $\alpha \! =\!0$ (or 2) the l.h.s.\ reduces to unity and one obtains the 
ghost DSE of standard Faddeev--Popov theory. For $\alpha\! =\!1$ both
terms on the r.h.s.\ are identical and add up to that of Eq.~(\ref{ghDSEs1}).

The main difference, as compared to ordinary Faddeev--Popov gauge,
in an explicit representation of the ghost DSE will be new type of diagrams
generated by the four-ghost interaction. 
The formal structure of the gluon DSE, on the other hand, remains unchanged. 

For completeness we have provided a derivation of the ghost DSE starting
directly from the Lagrangean (\ref{Lagrangian}) in appendix A. For all details
the interested reader is refered to this appendix as well as appendix B
which contains the definitions of Green's functions and the decompositions of 
full into connected  and one-particle irreducible Green's functions.
Employing the definitions of the bare ghost-gluon and the bare four-ghost
vertex, see appendix B, the Dyson--Schwinger equation for the ghost propagator
in coordinate space reads:
\beqa
[D_G^{ab}(x-y)]^{-1}
&=&\tilde{Z}_3 [D_G^{(0) ab}(x-y)]^{-1} \nonumber\\
&-& \tilde{Z}_1\int d^4zd^4ud^4vd^4z_1d^4z_2d^4z_3 \nonumber\\
&& \hspace*{1cm} \Gamma^{(0) bde}_\mu (y,u,v)  
D_{\mu \nu}^{ef}(v-z_1) \Gamma_\nu^{fha}(z_1,z_3,x)D_G^{hd}(u-z_3).
\nonumber\\  
&-& \tilde{Z}_4 \int d^4ud^4v \Gamma_{4gh}^{(0) bdfa}(x,u,v,y)  D_G^{fd}(v-u) \nonumber\\
&-& \tilde{Z}_4 \frac{1}{2} \int d^4zd^4ud^4vd^4u_1d^4u_2d^4u_3d^4u_4 
\Gamma^{(0) bdgf}_{4gh}(y,z,v,u) D_G^{fe}(u-u_4)\nonumber\\
&& \hspace*{1cm} \times
D_G^{gi}(v-u_2) \Gamma_{4gh}^{jaei}(u_3,x,u_4,u_2) D_G^{jd}(u_3-z) \nonumber\\ 
&-& \tilde{Z}_4 \frac{1}{2} \int d^4zd^4ud^4vd^4u_1d^4u_2d^4u_3d^4u_4d^4u_5 \nonumber\\
&& \hspace*{1cm} \Gamma^{(0) bdgf}_{4gh}(y,z,v,u) 
D_{\mu \nu}^{ek}(u_1-u_4) D_G^{fl}(u-u_5) \nonumber\\
&&  \hspace*{1cm} \times \Gamma^{kal}_\nu(u_4,x,u_5) D_G^{gi}(v-u_2) 
\Gamma^{eij}_\mu(u_1,u_3,u_2) D_G^{jd}(u_3-z). \nonumber\\ 
\label{DSEe_main}
\eeqa
Fourier transformation to momentum space yields:
\beqa
[D_G(p)]^{-1}
&=& \tilde{Z}_3 [D_G^{(0)}(p)]^{-1} \nonumber\\ 
&-& \tilde{Z}_1 \frac{g^2 N_c}{(2\pi)^4}\int d^4q \:\Gamma^{(0)}_\mu (p,q) \: 
D_{\mu \nu}(p-q) \:\Gamma_\nu(q,p)D_G(q) \nonumber\\ 
&-& \tilde{Z}_4 \frac{g^2 N_c}{(2\pi)^4} \int d^4q \: 
\Gamma_{4gh}^{(0)} \:  D_G(q) \nonumber\\
&+& \tilde{Z}_4 \frac{1}{2} \frac{g^4 N_c^2}{(2\pi)^8}\int d^4q_1q_2 \Gamma^{(0)}_{4gh} \:
D_G(q_1) \:D_G(p-q_1-q_2) \:\Gamma_{4gh}(p,q_1,q_2)\: D_G(q_2) 
\nonumber\\ 
&-& \tilde{Z}_4 \frac{1}{4} \frac{g^4 N_c^2}{(2\pi)^8} \int d^4q_1q_2 \: \Gamma^{(0)}_{4gh}\:  
D_{\mu \nu}(p-q_1) \: D_G(q1) \nonumber\\
&&  \hspace*{1cm} \times \Gamma_\nu(p,q_1)\:  D_G(q_2)\: 
\Gamma_\mu(-p+q_1+q_2,q_2)\: D_G(p-q_1-q_2). \nonumber\\ 
\label{DSE-ghost-ren}
\eeqa
The color traces have already been carried out and the reduced vertices defined in appendix B have been used.
The four-ghost interaction generates three new diagrams in the ghost equation, a tadpole contribution and two
two-loop diagrams. Furthermore the bare ghost-gluon vertex depends on the gauge parameter $\alpha$,
\beqa
\Gamma_\mu^{(0) abc}(k,p,q) &=& g f^{abc} (2 \pi)^4 \delta^4(k+q-p) \Gamma_\mu^{(0)}(p,q) \nonumber\\
\Gamma_\mu^{(0)}(p,q) &=& \left[ \left(1-\frac{\alpha}{2}\right)q_\mu + \frac{\alpha}{2} p_\mu \right].
\eeqa
Note the symmetry between the ghost momentum $p_\mu$ and the
antighost momentum $q_\mu$, when the gauge parameter $\alpha$ is set to one. 

\subsection{Projection of the gluon equation}

The respective equation for the gluon propagator is formally the same as in the Faddeev-Popov case.
Differences occur in the explicit form of the bare ghost-gluon vertex and the dressed vertices  
in general depend on the gauge parameters. The gluon DSE reads
\beqa
[D(p)]^{-1}_{\mu \nu}
&=& Z_3 [D^{(0)}(p)]^{-1}_{\mu \nu} \nonumber\\ 
&+& \tilde{Z}_1 \frac{g^2 N_c}{(2\pi)^4}\int d^4q \:\Gamma^{(0)}_\mu (p,q) \: 
D_G(p-q) \:\Gamma_\nu(q,p)D_G(q)
\nonumber\\ 
&-& Z_1 \frac{1}{2}\frac{g^2 N_c}{(2\pi)^4}\int d^4q \:\Gamma^{(0)}_{\mu \rho \sigma} (p,q) \: 
D_{\rho \rho^\prime}(p-q) \:\Gamma_{\rho^\prime \nu \sigma^\prime}(q,p)D_{\sigma \sigma^\prime}(q) \nonumber\\
&-& Z_4  \frac{1}{2} \frac{g^2 N_c}{(2\pi)^4} \int d^4q \: \Gamma^{(0)}_{\mu \nu \rho \sigma}
 \:  D_{\rho \sigma}(q) \nonumber\\
&-&  Z_4 \frac{1}{6} \frac{g^4 N_c^2}{(2\pi)^8}\int d^4q_1q_2 
\Gamma^{(0)}_{\mu \rho \sigma \lambda} \:
D_{\rho \rho^\prime}(q_2) \:D_{\sigma \sigma^\prime}(p-q_2-q_1) 
\:\Gamma_{\rho^\prime \nu \lambda^\prime \sigma^\prime}(p,q_1,q_2)\: 
D_{\lambda \lambda^\prime}(q_1) \nonumber\\ 
&-& Z_4  \frac{1}{2} \frac{g^4 N_c^2}{(2\pi)^8} \int d^4q_1q_2 \: 
\Gamma^{(0)}_{\mu \rho \sigma \lambda}\:  
D_{\rho \rho^\prime}(p-q_1-q_2) \: D_{\sigma \sigma^\prime}(q_2) \nonumber\\
&&  \hspace*{1cm} \times \Gamma_{\rho^\prime \zeta \sigma^\prime}(p-q_1-q_2,q_2)\:  
D_{\zeta \zeta^\prime}(p-q_1)\: \Gamma_{\zeta^\prime \nu \lambda^\prime}(p-q_1,q_1)
\: D_{\lambda \lambda^\prime}(q_1). \nonumber\\ 
\label{DSE-gluon-ren}
\eeqa

\begin{figure}
\epsfig{file=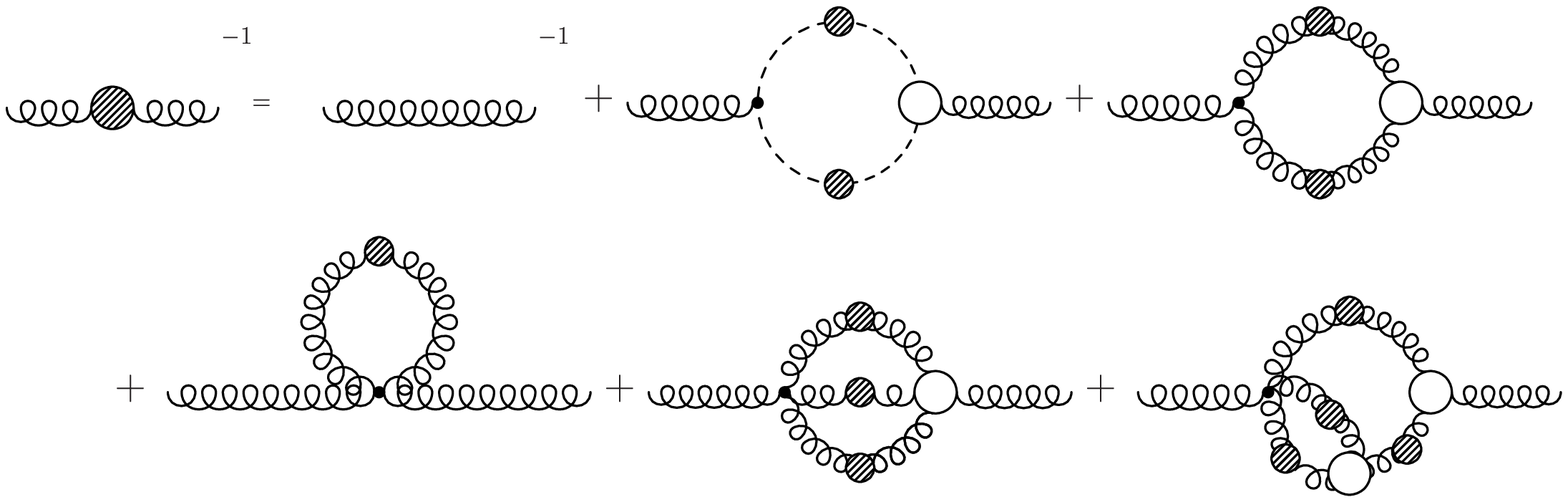,width=10cm}
\epsfig{file=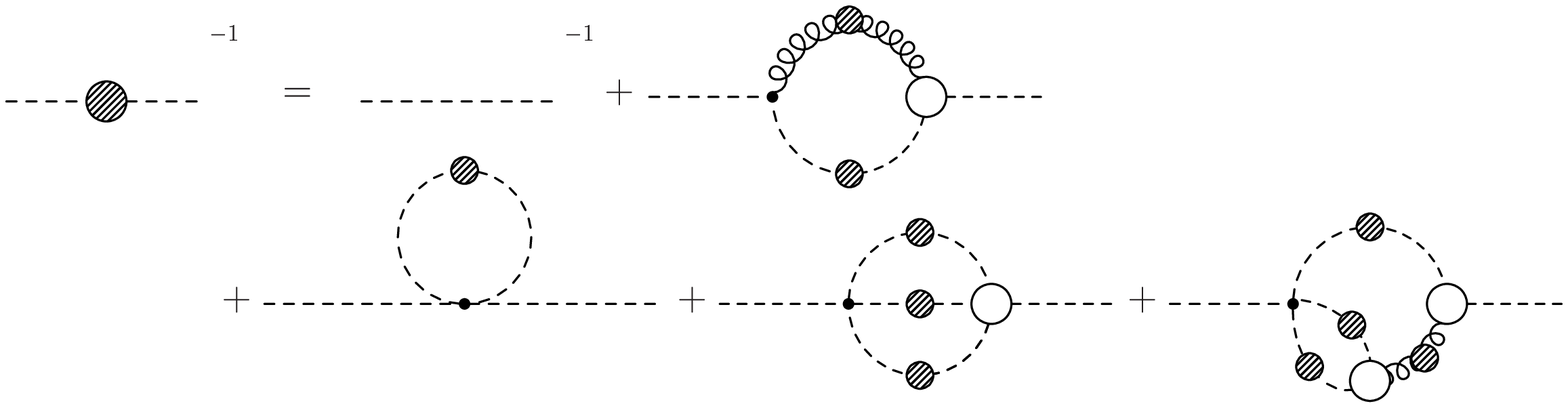,width=10cm}
\caption{The coupled gluon and ghost Dyson--Schwinger equations from a BRS and Anti-BRS symmetric
Lagrangean. Each equation contains one-loop diagrams, a tadpole contribution and a sunset and a squint diagram.}
\label{DSEpic}
\end{figure}
Both equations are shown diagrammatically in fig.\ \ref{DSEpic}.
One clearly sees the striking similarity between the ghost and the gluon equation once a 
four-ghost interaction has been introduced. Both equations have bare and one
loop parts, a tadpole contribution, 
a sunset and a squint diagram.

\begin{figure}
\begin{center}
\epsfig{file=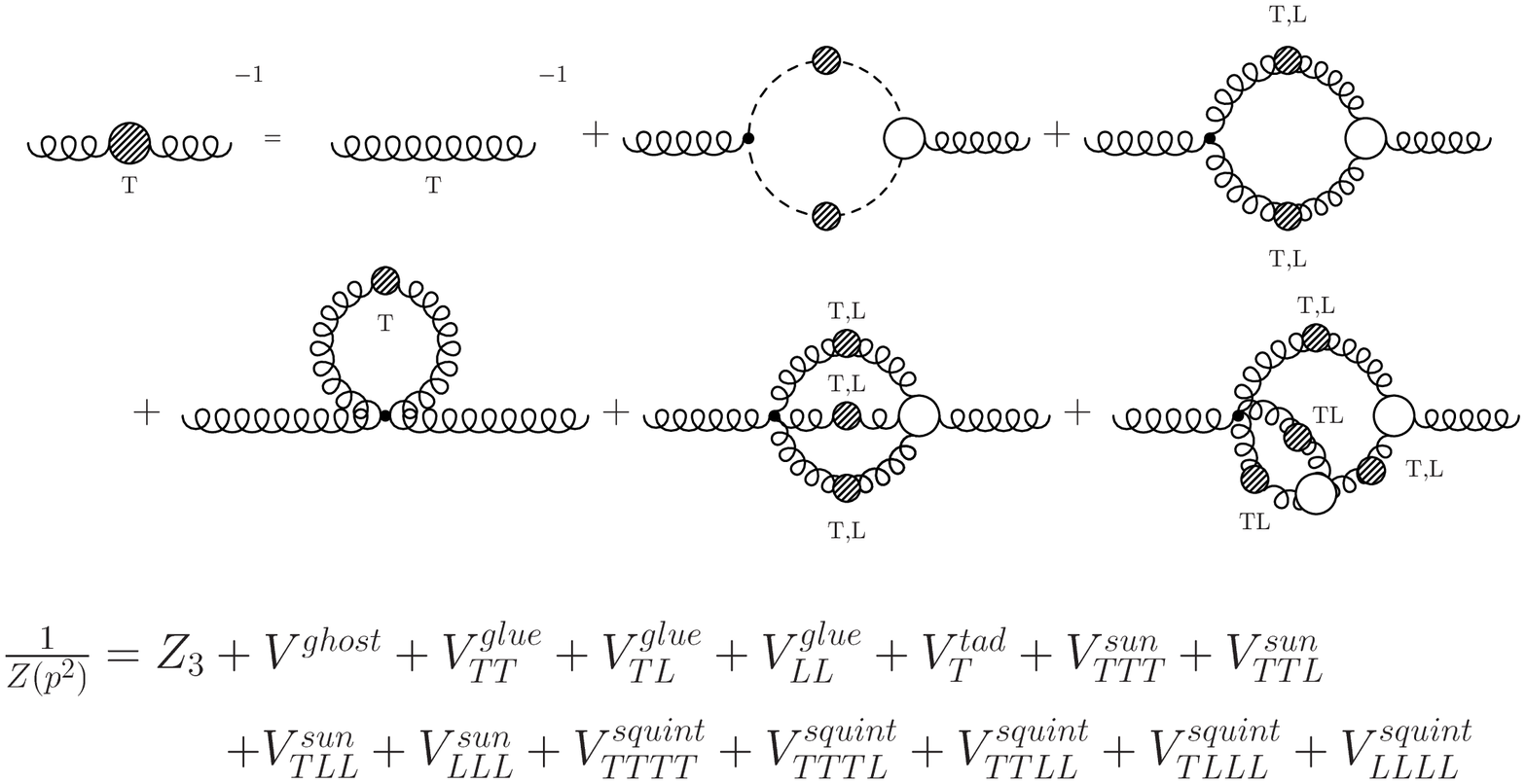,width=145mm}
\epsfig{file=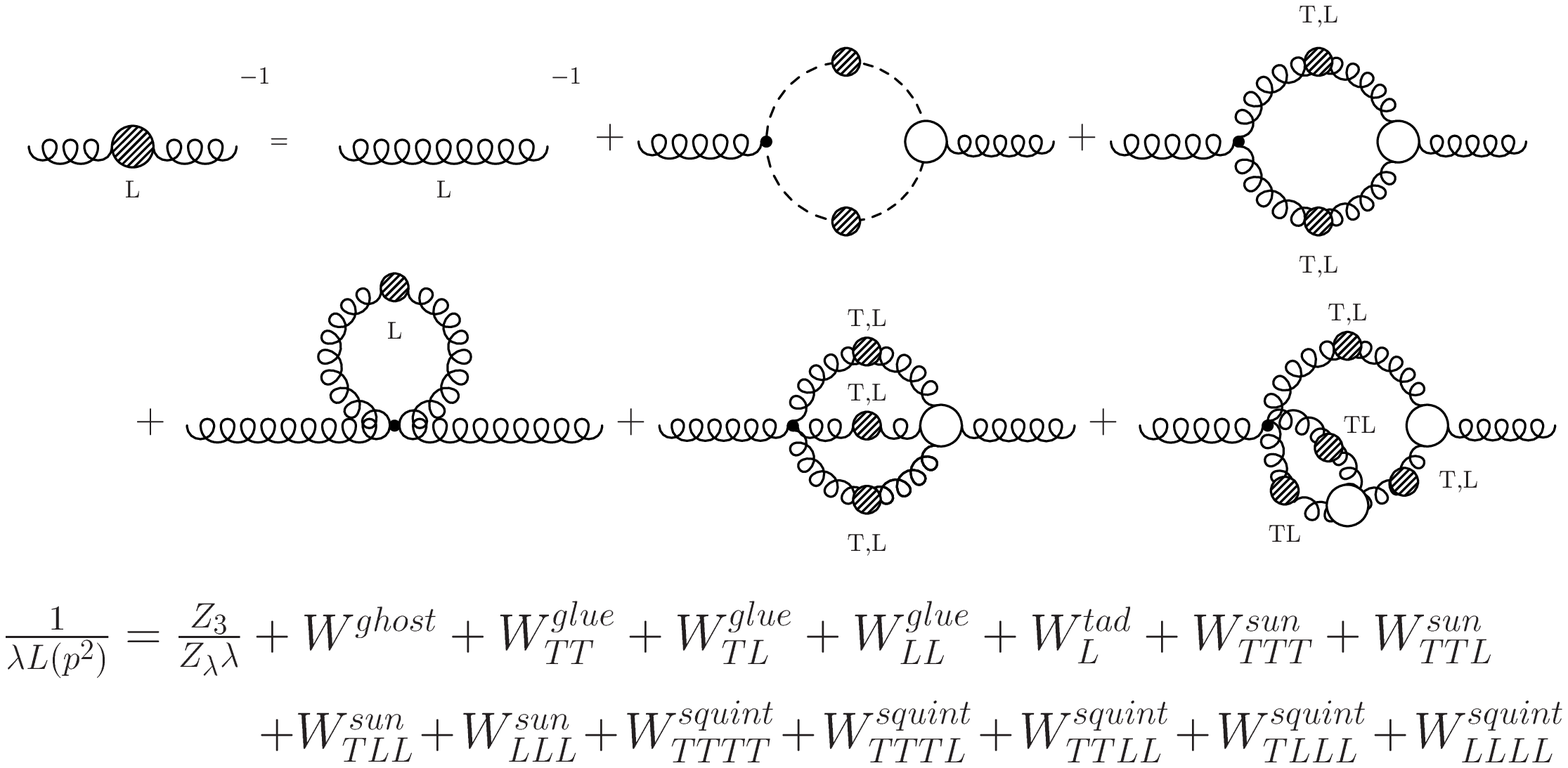,width=145mm}
\epsfig{file=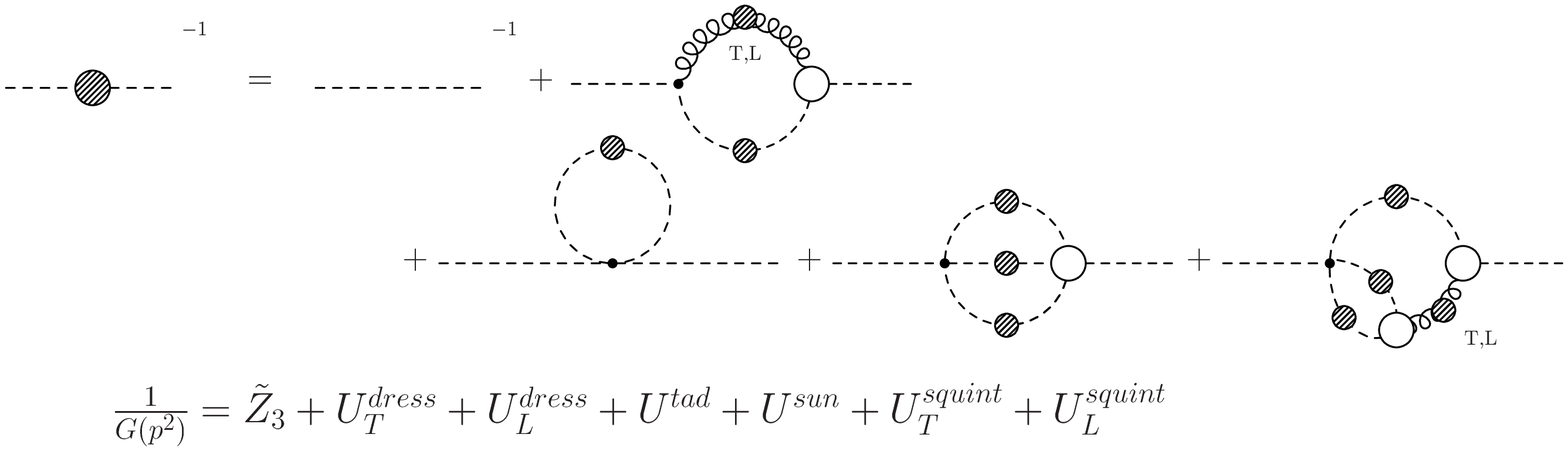,width=145mm}
\end{center}
\caption{Various contributions from the respective diagrams in the 
transverse and longitudinal gluon equation 
and the equation for the ghost dressing function.
}
\label{DSE_proj}
\end{figure}
In order to sort the various contributions of the gluon equation to the inverse of the
gluon propagator on the left hand side we project the equation on its longitudinal and transverse
parts. It is well known that for linear covariant gauges, $\alpha=0$, 
the longitudinal part of the gluon propagator remains undressed 
\cite{Alkofer:2001wg}. However, away from linear covariant gauges
this is not the case as can be seen from the corresponding Slavnov-Taylor identity derived in 
\cite{Baulieu:1982sb}. We then have three dressing functions in the general case and the propagators are given by
\beqa
D_{\mu \nu}(p) &=& \left[D_{\mu \nu}(p)\right]_T + \left[D_{\mu \nu}(p)\right]_L \nonumber\\
&=& \left( \delta_{\mu \nu} - \frac{p_\mu p_\nu}{p^2}\right)
\frac{Z(p^2)}{p^2} + \lambda L(p^2) \frac{p_\mu p_\nu}{p^4}, \\
D_G(p) &=& -\frac{G(p)}{p^2}. 
\eeqa
The transversal and longitudinal gluon dressing functions $Z(p^2)$ and $L(p^2)$ can be extracted by contracting
the gluon equation with the transversal and longitudinal projector respectively. The results are given
graphically in fig.\ \ref{DSE_proj}, where we also specify our notation for the different contributions
being analyzed in the next section. Contributions in the transversal part of the gluon equation are denoted by
the symbol $V$, contributions in the longitudinal part by $W$ and the ones in the ghost equation by $U$.
The subscripts $T$ and $L$ indicate the respective parts of the gluon propagator running around in the loops
of the diagrams and abbreviations for the diagrams are used. For example the symbol $W^{sun}_{LLT}$ denotes
a contribution from the sunset diagram to the longitudinal gluon equation with two longitudinal and one 
transverse part of the gluon propagator running in the loop. To isolate the dressing functions the left
hand side of the equations have already been divided by factors of $3 p^2$ and $p^2$ respectively. 

\subsection{Generalized Slavnov--Taylor Identities \label{STI}}

To derive the generalization of the Slavnov--Taylor identity for the
longitudinal gluon propagator we start from the following BRS variations:
\begin{eqnarray}
  \label{glBRS}
  s_r \big( \partial A^a_x \bar c^b_y \big) &=&
          - \widetilde Z_3 \, (\partial D_r c)^a_x \, \bar c^b_y \, +\,
     \partial A^a_x \, \big(   B - \widetilde Z_1  \, 
     \frac \alpha 2 \, (\bar c \times c)     \big)^b_y
\; ,\\
  \bar s_r \big( \partial A^a_x  c^b_y \big) &=&
         - \widetilde Z_3 \, (\partial D_r \bar c)^a_x \, c^b_y \, - \,
     \partial A^a_x \, \big(   B + \widetilde Z_1  \,(1-\frac \alpha 2) \,
(\bar c \times c)  \big)^b_y     \; . \label{glBRS2}
\end{eqnarray}
The corresponding vacuum expectation values vanish, and 
taking combinations of the expectation values of these equations we obtain:
\begin{eqnarray}
  \label{glBRS1}
  0 &=&  (1-\frac \alpha 2) \, \langle \, s_r 
  \big( \partial A^a_x \bar c^b_y \big)
                      \, \rangle  \, - \,
        \frac \alpha 2 \,  \langle \, \bar s_r \big( \partial A^a_x  c^b_y \big)
                       \, \rangle 
\nonumber \\
    &=&   - \widetilde Z_3 \,
        (1-\frac \alpha 2) \, \langle  \,  
	(\partial D_r c)^a_x \, \bar c^b_y\,
  \rangle \,  + \,  \widetilde Z_3 \, \frac \alpha 2\,  \langle  \,  
  (\partial D_r \bar c)^a_x \,  c^b_y \, \rangle \, + \, \langle\,
                  \partial A^a_x \,   B^b_y \, \rangle \; .
\end{eqnarray}
Upon insertion of the e.o.m.\ of the $B$-field, 
$ Z_\lambda \lambda \, B \,=\, i Z_3 \partial A$,  
this directly leads to Eq.~(\ref{ghDSEs3}).
On the other hand, the ghost DSEs from Eqs.~(\ref{ghDer}) 
and (\ref{aghDer}) allow to eliminate the first two terms on
the r.h.s., multiplying to them appropriate factors of $\frac \alpha 2$ and 
$1-\frac \alpha 2$ and inserting these expression in eq.\ (\ref{glBRS1})
yields
\begin{eqnarray}
  \label{glSTI}
   {Z_3} \,  \langle \,
      \partial A^a_x \, \partial A^b_y \, \rangle  &=&  
      {Z_\lambda \, \lambda} \,
\Big\{ \, \delta^{ab} \delta_{xy}   \\
 & - & \, i \widetilde Z_1  \, \frac \alpha 2 (1-\frac \alpha 2) \, 
 \langle \, (\partial A
      \times  c)^a_x \, \bar c^b_y \, \rangle \, + \, 
      \frac{Z_\lambda \, \lambda}{ Z_3}
      \, \widetilde Z_1^2 \, \frac \alpha 2 (1-\frac \alpha 2)^2 \,
  \langle \, \big( (\bar c \times  c ) \times c \big)^a_x \, \bar c^b_y \,
                      \rangle  \nonumber\\
 & + &\, i \widetilde Z_1  \, \frac \alpha 2 (1-\frac \alpha 2) \, \langle \,
      (\partial A\times \bar c)^a_x \,  c^b_y \, \rangle \,
     + \, \frac{Z_\lambda \, \lambda}{Z_3}
      \, \widetilde Z_1^2 \, \frac { \alpha^2} 4 (1-\frac \alpha 2) \,
    \langle \,   \big( (\bar c \times  c ) \times \bar c \big)^a_x \,
           c^b_y \,  \rangle \, \Big\}\; .  \nonumber
\end{eqnarray}
This generalizes the Slavnov--Taylor identity for the longitudinal part of the
gluon propagator which, contrary to the standard Faddeev--Popov gauges, does in
general acquire renormalization by the interactions, {\it c.f.\/}
Eq.~(\ref{Zxi}). On the r.h.s.\ of the Slavnov--Taylor identity, 
the terms in the 3rd line are the Faddeev--Popov
conjugate of those in the 2nd. In the ghost anti-ghost symmetric case for
$\alpha \! =\! 1$ they are identical.  In this case the Slavnov--Taylor identity
simplifies,
\begin{equation}
  \label{glSTIs}
 {Z_3} \,  \langle \,
      \partial A^a_x \, \partial A^b_y \, \rangle  \,=\, {Z_\lambda \, \lambda} \,
\Big\{ \, \delta^{ab} \delta_{xy}   \, - \, i \widetilde Z_1 \, \frac{1}{2}
 \, \langle \, (\partial A \times  c)^a_x \, \bar c^b_y \, \rangle \, + \,
             \frac{Z_\lambda \, \lambda}{ Z_3}
      \,  \frac{{\widetilde Z}_1^2 }{4}  \,
  \langle \, \big( (\bar c \times  c ) \times c \big)^a_x \, \bar c^b_y \,
                      \rangle \, \Big\}  \; .
\end{equation}
Note that close to the Landau gauge the corrections to the unity of the
standard Faddeev--Popov gauges are suppressed by one order in the gauge
parameter $\lambda$.

A further Slavnov--Taylor identity is obtained by adding the 
expectation values of the
BRS variations in Eqs.~(\ref{glBRS}) and (\ref{glBRS2}):
\begin{eqnarray}
  \label{glSTIa}
  0 &=&  \langle \, s_r \big( \partial A^a_x \bar c^b_y \big)
                      \, \rangle  \, + \,
           \langle \, \bar s_r \big( \partial A^a_x  c^b_y \big)
                       \, \rangle 
\nonumber \\
    &=&   - \widetilde Z_3 \,
        \langle  \,  (\partial D_r c)^a_x \, \bar c^b_y\,
  \rangle \,  - \,  \widetilde Z_3 \,  \langle  \,  (\partial D_r \bar
                      c)^a_x \,  c^b_y \,  \rangle \, -  \,
              \widetilde Z_1 \,  \langle\,
                  \partial A^a_x \,  (\bar c \times c)^b_y \, \rangle \; .
\end{eqnarray}
This leads to Eq.~(\ref{auxSTI}) as promised in the previous subsection.

These Slavnov--Taylor identities indicate that the Landau gauge limit $\lambda
\to 0$ is smooth. Based on Eqs.~(\ref{glSTI}) and (\ref{glSTIa})  one may
anticipate that an infrared massless-like longitudinal part of the gluon
propagator leads for sufficiently small values of the gauge parameter
$\lambda$ to the same infrared  enhancement of ghosts as observed in
the Landau gauge.

\section{Infrared analysis with bare vertices for arbitrary gauge para\-meters}

In this section we will analyse the behaviour of the two-point functions at
small momenta~$p^2$. 
We will employ a truncation scheme that successfully has been applied in the case of Landau gauge 
\cite{Lerche:2002ep,Zwanziger:2001kw,Fischer:2002hn} and
explore its applicability to general gauges. 

An interesting result of the investigations in Landau gauge is the observation, that there is no
qualitative difference of the solutions found with bare vertices or with vertices dressed
by the use of Slavnov--Taylor identities. This has not only been found in truncations using angular approximations
\cite{vonSmekal:1997is,Atkinson:1998tu} for the integrals, but has been confirmed recently for a 
range of possible vertex dressings in a truncation scheme without any angular approximations \cite{Lerche:2002ep}.
The reason for
this somewhat surprising result has been attributed to the non-renormalization of the ghost-gluon vertex in
Landau gauge, that is $\tilde{Z}_1=1$. It seems as if the violation of gauge invariance using a bare vertex is
not that severe in Landau gauge such that the resulting equations still provide meaningful results.
In the following we will explore to what extent such a simple truncation idea is applicable in
other gauges where $\tilde{Z}_1 \not= 1$.

In Landau gauge the coupled set of Dyson-Schwinger equations is solved by pure power
laws for the ghost and gluon dressing functions. Such solutions are determined analytically by plugging a
power law {\it ansatz} in the equations and match appropriate powers on the left and right hand side. 
Once several power solutions have been found the remaining task is to single out the one matching 
the numerical solution of the renormalized equation. 
In Landau gauge it has been shown that indeed only one of the power solutions found in
refs.\ \cite{Lerche:2002ep,Zwanziger:2001kw} is the correct infrared limit of the 
renormalized solution \cite{Fischer:2002hn} by solving the equations numerically for all momenta.
In the following we will investigate whether there are power solutions at all using bare vertices for general gauge
parameter $\alpha$ and $\lambda$.

Now we employ the power law ansatz for the dressing functions,
\beq
G(x) = Bx^\beta, \:\:\:\:
Z(x) = Ax^\sigma, \:\:\:\:
L(x) = Cx^\rho,
\eeq 
where $x=p^2$ has been used. Together with the expressions for the bare vertices given in appendix B
we plug the power laws into the ghost and the gluon equation. The formulae for the various integrals 
are given in appendix C. The straightforward but tedious algebra is done
with the help of the algebraic manipulation program FORM \cite{Form}. 
In ref.~\cite{Lerche:2002ep}
it has been shown that the renormalization functions $Z_3$ and $\tilde{Z}_3$ do not play a role in the 
determination of possible power solutions of the equations in the infrared region of momentum. 
Furthermore the tadpoles just give constant 
contributions to the respective propagators which vanish in the process of renormalization. 
Thus we safely omit them in the present investigation.

For the most general gauges, $\alpha \ne 0$ and $\lambda \ne 0$, we obtain the following structure:
\beqa
B^{-1}x^{-\beta} &=& x^{\sigma+\beta}(U^\prime)^{dress}_{T} + x^{\rho+\beta}(U^\prime)^{dress}_{L} 
  + x^{3\beta}(U^\prime)^{sun} \nonumber\\
&&+x^{\sigma+3\beta} (U^\prime)^{squint}_{T} +x^{\rho+3\beta}  (U^\prime)^{squint}_{L} \label{gheq1}\\
A^{-1}x^{-\sigma} &=& x^{2\beta}(V^\prime)^{ghost} +x^{2\sigma} (V^\prime)^{glue}_{TT} 
  + x^{\sigma+\rho}(V^\prime)^{glue}_{TL} + x^{2\rho}(V^\prime)^{glue}_{LL} \nonumber\\
&&+ x^{3\sigma}(V^\prime)^{sun}_{TTT} + x^{2\sigma+\rho}(V^\prime)^{sun}_{TTL}
  + x^{\sigma+2\rho}(V^\prime)^{sun}_{TLL}+ x^{3\rho}(V^\prime)^{sun}_{LLL}\nonumber\\
&&+ x^{4\sigma}(V^\prime)^{squint}_{TTTT} + x^{3\sigma+\rho}(V^\prime)^{squint}_{TTTL}
  + x^{2\sigma+2\rho}(V^\prime)^{squint}_{TTLL} \nonumber\\
&&+ x^{\sigma+3\rho}(V^\prime)^{squint}_{TLLL}
  + x^{4\rho}(V^\prime)^{squint}_{LLLL} \label{gl_trans1}\\
\left(C \lambda\right)^{-1} x^{-\rho}
&=& + x^{2\beta}(W^\prime)^{ghost} + x^{2\sigma}(W^\prime)^{glue}_{TT} 
  + x^{\sigma+\rho}(W^\prime)^{glue}_{TL}  \nonumber\\
&&+ x^{3\sigma}(W^\prime)^{sun}_{TTT} + x^{2\sigma+\rho}(W^\prime)^{sun}_{TTL}
  + x^{\sigma+2\rho}(W^\prime)^{sun}_{TLL}+ x^{3\rho}(W^\prime)^{sun}_{LLL}\nonumber\\
&&+ x^{4\sigma}(W^\prime)^{squint}_{TTTT} + x^{3\sigma+\rho}(W^\prime)^{squint}_{TTTL}
  + x^{2\sigma+2\rho}(W^\prime)^{squint}_{TTLL} \nonumber\\
&&+ x^{\sigma+3\rho}(W^\prime)^{squint}_{TLLL}. \label{gl_long1}
\eeqa
Here the primed quantities are momentum independent 
functions of $\beta$, $\sigma$ and $\rho$, {\it c.f.\/} fig.~\ref{DSE_proj} where the corresponding
unprimed, momentum dependent quantities have been introduced. 
The pattern of the equation is such that each primed factor on the right hand side is
accompanied by the squared momentum $x$ to the power of the dressing
function content of the respective diagram. 
In appendix D we demonstrate how such a pattern emerges for
example from  
the sunset diagram in the ghost equation, $(U)^{sun}$. Note that the contributions $(W^\prime)^{glue}_{LL}$
and $(W^\prime)^{squint}_{LLLL}$ are zero and therefore missing
in the longitudinal gluon equation (\ref{gl_long1}) as momentum conservation cannot hold with three
longitudinal gluons in the three gluon vertex. 

For the following argument we focus on one particular contribution on each right hand side of the equations:
\beqa
B^{-1}x^{-\beta} &=&  x^{3\beta}(U^\prime)^{sun} + \ldots \label{gheq2}\\
A^{-1}x^{-\sigma} &=& x^{4\sigma}(V^\prime)^{squint}_{TTTT} + \ldots \label{gl_trans2} \\
\left(C \lambda\right)^{-1} x^{-\rho}
&=& x^{3\rho}(W^\prime)^{sun}_{LLL} + \ldots \label{gl_long2}
\eeqa

The coefficients $(U^\prime)^{sun}$, $(V^\prime)^{squint}_{TTTT}$ and $(W^\prime)^{sun}_{LLL}$ are nonzero
and explicitly given in appendix D. 
{\it First} it is now easy to see from equations (\ref{gheq2}), (\ref{gl_trans2}) and (\ref{gl_long2}) that neither 
$\beta$ nor $\sigma$ nor $\rho$ can be negative. If one of these powers would be negative the
limit $x \rightarrow 0$ would lead
to a vanishing left hand side of the respective equation whereas the right hand side is singular in this limit.
This is a contradiction as the power on the left hand side of the equation should match the leading power on
the right hand side.  
{\it Second} if one of $\beta$, $\sigma$ or $\rho$ would be positive, then the diverging left hand side
of the respective equation would require a diverging counterpart on the right hand side.
However, all powers on the right hand side are positive as we already concluded that
$\beta$, $\sigma$ or $\rho$ are not negative and there are no minus sines in any powers on the right
hand sides, {\it c.f.\/} Eqs.~(\ref{gheq1}), (\ref{gl_trans1}) 
and (\ref{gl_long1})). Therefore for positive powers all terms on the right hand side vanish in the limit
$x \rightarrow 0$ which leads again to a contradiction.
The last possibility is then $\beta=\sigma=\rho=0$, but then one gets perturbative logarithms
on the right hand side of the equation which do not match the constant
on the left hand side.
Thus in the all-bare-vertex truncation there is no 
power solution for general values of the gauge parameters $\lambda \not =0$
and $\alpha \not=0$. Based on the considerations on the Slavnov--Taylor
identities given in the previous section we therefore arrive at the conclusion 
that this truncation is
insufficient to determine the infrared behaviour of the propagators even
qualitatively.   

There are two limits for the gauge parameters $\alpha$ and $\lambda$ in which the situation changes.
The first one is $\alpha=0$, that is ordinary linear covariant gauges. Due to the corresponding
Slavnov--Taylor identity the
longitudinal part of the gluon propagator remains undressed, $L(p^2)=1$ \cite{Alkofer:2001wg}.
However, replacing dressed vertices by bare ones in the infrared, this
identity might be violated (which does not happen in perturbation theory,
of course). We therefore employ the
general expression $L(p^2) = C (p^2)^\rho$ for the longitudinal gluon dressing function and explore
whether the limit $\rho \rightarrow 0$ can be taken with bare vertices. In the ghost equation 
the squint as well as the sunset diagram disappear and we are left with the one-loop contributions
$U^{dress}_T$ and $U^{dress}_L$. The explicit expression for the ghost equation is given by (cf. appendix D) 
\beqa
B^{-1}x^{-\beta} &=&  U^{dress}_T + U^{dress}_L \\
&=& x^{(\beta  + \sigma )} \frac {g^2 N_c \tilde{Z}_1 AB}{16 \pi^2} \, \frac {-3} 
{2\,(\beta  + \sigma )\,( - 1 + \beta  + \sigma )} 
\frac{\Gamma (2 + \beta ) \,\Gamma (1 + \sigma ) \,
\Gamma (2 - \beta  - \sigma 
)}{\Gamma (1 - \beta )\,\Gamma (2 - \sigma )\,\Gamma (3 + \beta 
 + \sigma )} \nonumber\\
&&-x^{(\beta+\rho)} \frac{g^2 N_c \tilde{Z}_1 \lambda B C  }{16 \pi^2}  
\:\: \frac{\rho + 1/2}{\beta} \:\:
\frac{\Gamma(2+\beta) \Gamma(1+\rho) \Gamma(-\beta-\rho)}
{\Gamma(-\beta) \Gamma(2-\rho) \Gamma(3+\rho+\beta)}  \,. \label{ghost3}
\eeqa
For $\rho \rightarrow 0$ we run into the same contradiction as explained above for general values of
the gauge parameters $\alpha$ and $\lambda$. However, admitting the generation of a (spurious) longitudinal 
gluon dressing this contradiction can be resolved in the following way: Equation (\ref{gl_trans2})
for the transversal gluon dressing function does not change in structure, therefore $\sigma > 0$. 
Furthermore we have $\rho > 0$ from eq.~(\ref{gl_long2}). Then we have $-2\beta=\sigma$ and/or $-2\beta=\rho$
in the ghost equation (\ref{ghost3}) and $\beta<0$, {\it i.e.}
a diverging ghost dressing function in the infrared. From this it follows
immediately, that the ghost loop is the dominant contribution in both, the equations for the transversal
and longitudinal gluon dressing function. From these two equations we therefore infer  
\beq
-\beta=\sigma/2=\rho/2=:\kappa \,,
\label{kappa}
\eeq
which is consistent with the ghost equation. We thus find an infrared vanishing gluon 
dressing and a singular ghost dressing function for all values of the gauge parameter $\lambda$.
This result is identical to the one in Landau gauge
\cite{Zwanziger:2001kw,Lerche:2002ep,Fischer:2002hn}.
However, a word of caution is in order. In Landau gauge there are indications
\cite{vonSmekal:1997is,Watson:2001yv} that the general result (\ref{kappa})
does not change when the vertices are dressed. This has been confirmed recently for a  
range of possible vertex dressings \cite{Lerche:2002ep}.
It is an as yet open question whether this is true for $\lambda \ne 0$
in the same way.

Having addressed the case of linear covariant gauges with $\alpha=0$ we now turn to 
the other interesting limit, that is $\lambda=0$, while $\alpha \ne 0$. It is easy to see, that the
$\alpha$-dependence of
the Lagrangean (\ref{Lagrangian}) can be eliminated in this case by partial integration using
the constraint $\partial A=0$. However, on the level of the DSEs with
bare vertices there remain spurious $\alpha$-dependent terms on the right hand side of the gluon equation. 
In the next section we will investigate the dependence of the Landau
gauge solution on these spurious $\alpha$-terms.

\section{Solutions in Landau gauge}

To assess the influence of the spurious $\alpha$-terms in Landau gauge
we use the truncation scheme developed in \cite{Fischer:2002hn}. There the two loop diagrams in
the gluon equation have been neglected as they are subleading in the perturbative regime and
ghost loop dominance has been assumed in the infrared. In order to obtain the correct one loop
behaviour of the ghost and gluon dressing functions the gluon loop has been modified by replacing the
renormalization constant $Z_1$ by a momentum dependent function $\mathcal{Z}_1$:
\begin{equation} 
Z_1(L,s) \to {\cal Z}_1(x,y,z;s,L)  = 
\frac{G(y)^{(-2-6\delta)}}{Z(y)^{(1+3\delta)}}
\frac{G(z)^{(-2-6\delta)}}{Z(z)^{(1+3\delta)}}
\end{equation}
Here $L=\Lambda^2$ denotes a cutoff and $s=\mu^2$ a renormalization scale in units of squared momenta.
The momentum $x=p^2$ is the one flowing into the loop, $y=q^2$ is the loop momentum over which is
integrated and $z:=k^2=(p-q)^2$. Furthermore the anomalous dimension $\delta$ of the ghost
dressing function has been used.
The gluon equation is contracted with the general tensor
\beq
{\mathcal P}^{(\zeta )}_{\mu\nu} (p) = \delta_{\mu\nu} - 
\zeta \frac{p_\mu p_\nu}{p^2} 
\, . 
\label{Paproj}
\eeq 
As a completely transversal gluon equation would be independent of the parameter $\zeta$ the use of the general
projector provides an opportunity to test for violations of transversality due to the truncation. 
For $\zeta \not= 4$ one
has to take care of spurious quadratic divergencies that have to be subtracted in the
kernel of the gluon equation.

The coupled set of equations for the ghost and gluon dressing functions then read as follows
\begin{eqnarray} 
\frac{1}{G(x)} &=& Z_3 - g^2N_c \int \frac{d^4q}{(2 \pi)^4}
\frac{K(x,y,z)}{xy}
G(y) Z(z) \; , \label{ghostbare} \\ 
\frac{1}{Z(x)} &=& \tilde{Z}_3 + g^2\frac{N_c}{3} 
\int \frac{d^4q}{(2 \pi)^4} \frac{M(x,y,z)}{xy} G(y) G(z) 
\nonumber\\
&& \qquad +
 g^2 \frac{N_c}{3} \int \frac{d^4q}{(2 \pi)^4} 
\frac{Q(x,y,z)}{xy} Z(y) Z(z) {\cal Z}_1(y,z) \; .
\label{gluonbare} 
\end{eqnarray} 
The kernels ordered with respect to powers of $z:=p^2=(k-q)^2$ have the form:
\begin{eqnarray}
K(x,y,z) &=& \frac{1}{z^2}\left(-\frac{(x-y)^2}{4}\right) + 
\frac{1}{z}\left(\frac{x+y}{2}\right)-\frac{1}{4}\\
M(x,y,z) &=& \frac{1}{z} \left( \frac{(\zeta-1)\alpha^2-(\zeta-1)2\alpha+\zeta-2}{4}x + 
\frac{y}{2} - \frac{\zeta}{4}\frac{y^2}{x}\right)
+\frac{1}{2} + \frac{\zeta}{2}\frac{y}{x} - \frac{\zeta}{4}\frac{z}{x} \\
Q(x,y,z) &=& \frac{1}{z^2} 
\left( \frac{1}{8}\frac{x^3}{y} + x^2 -\frac{19-\zeta}{8}xy + 
\frac{5-\zeta}{4}y^2
+\frac{\zeta}{8}\frac{y^3}{x} \right)\nonumber\\
&& +\frac{1}{z} \left( \frac{x^2}{y} - \frac{15+\zeta}{4}x-
\frac{17-\zeta}{4}y+\zeta\frac{y^2}{x}\right)\nonumber\\
&& - \left( \frac{19-\zeta}{8}\frac{x}{y}+\frac{17-\zeta}{4}+
\frac{9\zeta}{4}\frac{y}{x} \right) \nonumber\\
&& + z\left(\frac{\zeta}{x}+\frac{5-\zeta}{4y}\right) + z^2\frac{\zeta}{8xy} +\frac{5}{4}(4-\zeta).
\label{new_kernels}
\end{eqnarray}
\begin{figure}
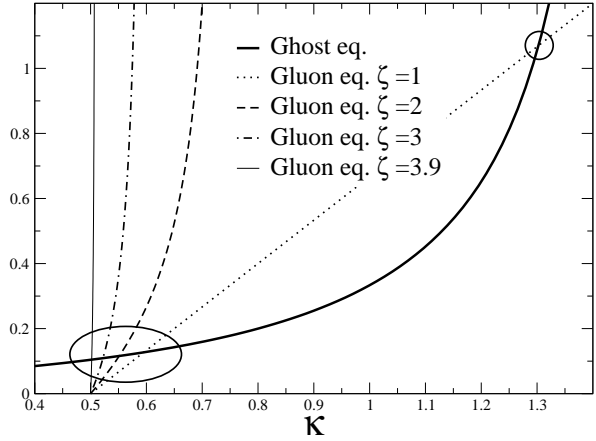

\centerline{
\epsfig{file=kappa.eps,height=6.5cm} \hspace*{0.7cm}
\epsfig{file=kappa2.eps,height=6.5cm}
}
\caption{\label{kappa.dat}
Here the graphical solution to equation (\ref{kappa2}) is shown. 
The thick line represents the
left hand side of equation (\ref{kappa2}), whereas the other curves depict 
the right hand side for different values of the parameters $\zeta$. The left figure shows results for
$\alpha=0$ and $\alpha=2$, whereas in the figure on the right $\alpha=1$. The ellipse marks the 
bulk of solutions between
$\kappa=0.5$ and $\kappa=0.6$ for $\zeta=1$, whereas the circles in the left figure show the 
movement of the solution for the Brown-Pennington case $\zeta=4$ from $\kappa=1$ to $\kappa=1.3$.
}
\end{figure}

First we accomplish the infrared analysis. With equation (\ref{kappa}) we employ the ansatz
\beq
Z(x) = Ax^{2 \kappa} \:\:\:\:\:\:\: G(x) = Bx^{-\kappa}
\eeq
in the equations (\ref{ghostbare}) and (\ref{gluonbare}). After integration we match coefficients of equal 
powers on both side of the equations and obtain
\begin{eqnarray} 
&&\frac{1}{18}
\frac{(2+\kappa)(1+\kappa)}{(3-2\kappa)}
    \nonumber\\
&&=\frac{(4\kappa-2)\,( - 1 + \kappa)}{(\zeta  - 1)\left[4\,\kappa
^{2}\,(\alpha ^{2} - 2\,\alpha  + 1) + 8\,\kappa\,\alpha
 \,(2 - \alpha )+ 3\,\alpha \,(\alpha  - 2)\right] + \kappa\,(10 - 7\,
\zeta )  - 6 + 3\,\zeta }
\; . \nonumber\\
\label{kappa2}
\end{eqnarray} 
The values of $\kappa$ for different projectors $\mathcal{P}^{(\zeta)}$ can be
read off fig.~\ref{kappa.dat}.
The curve given by the fully drawn line represents the term on the left hand side of equation (\ref{kappa2}), whereas
the other lines depict the right hand side for several values of the parameter $\zeta$.
Only the two $\zeta = 1$ solutions are manifestly independent of
$\alpha$, as pointed out in \cite{Lerche:2002ep}. The spurious
$\alpha$-dependence of the $\zeta = 4$ values reported therein,
here implies that general $\zeta$ solutions must necessarily show such
an $\alpha$-dependence also, whenever $\zeta \not= 1$.
However, the bulk of solutions between $\kappa=0.5$ and $\kappa=0.6$
remains nearly unchanged when $\alpha$ is varied, whereas most of the
solutions for $\kappa \ge 1$ disappear. 
For the Brown-Pennington projector $\zeta=4$ no solution can be found
for the symmetric case, $\alpha=1$, in complete agreement with the
findings of ref. \cite{Lerche:2002ep}. Indeed  
it has been shown \cite{Fischer:2002hn} that only the smaller solutions 
are those that connect to numerical results for finite momenta.

\begin{figure}[t]
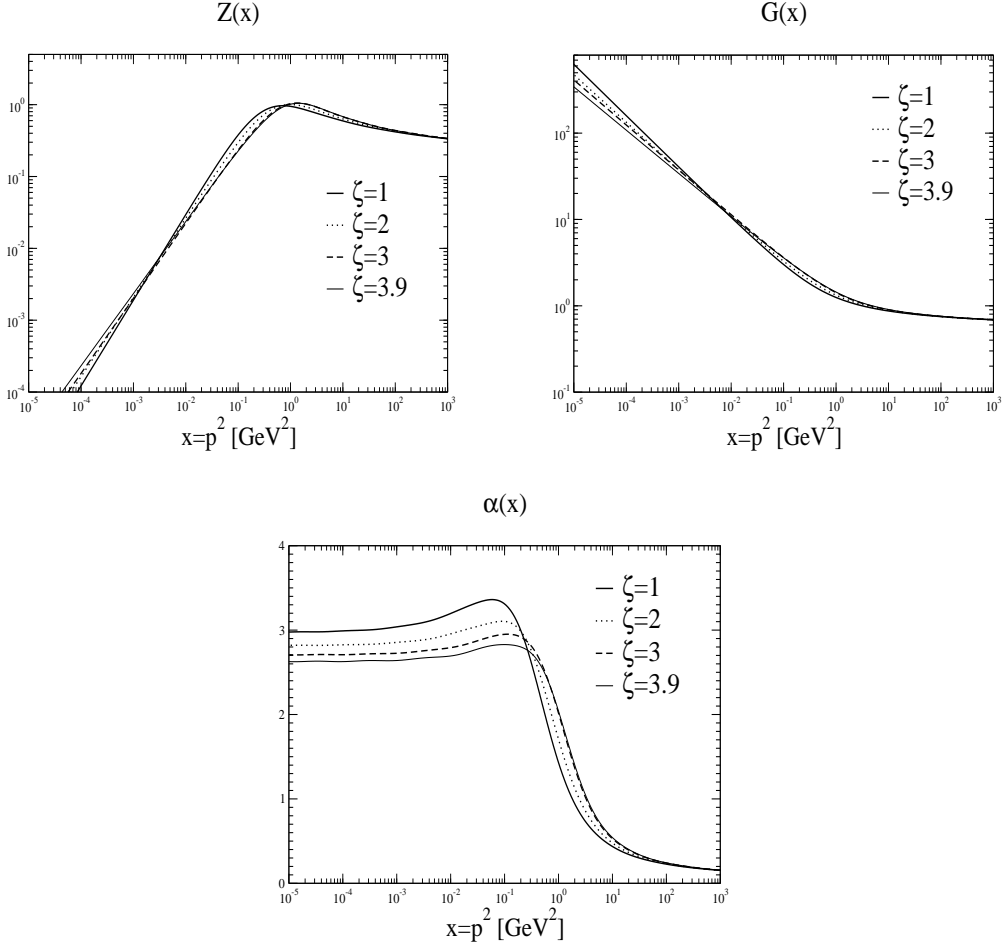

\vspace{-0cm}
\centerline{
\epsfig{file=z_symm.eps,width=6.0cm,height=6cm}
\hspace{1cm}
\epsfig{file=g_symm.eps,width=6.0cm,height=6cm}
}
\vspace{0.5cm}
\centerline{
\epsfig{file=a_symm.eps,width=6.0cm,height=6cm}
}
\caption{\label{new.dat}
Shown are the gluon dressing function, the ghost dressing function and
the running coupling in the truncation scheme \cite{Fischer:2002hn} for the gauge parameters 
$\alpha=1$ and $\lambda=0$ and different projectors $\mathcal{P}^{(\zeta)}$.}
\end{figure}

We now explore the impact of the spurious $\alpha$ term on the behaviour of the solutions
for all momenta $x$. We have solved eqs.~(\ref{ghostbare}) and (\ref{gluonbare}) numerically 
using the same technique as described in \cite{Fischer:2002hn}. The results can
be seen in fig.~\ref{new.dat}. As the dependence of the kernel of the ghost loop on $\alpha$ vanishes in the case of the
transverse projector, $\zeta=1$, this solution is the same as the one already calculated in \cite{Fischer:2002hn}.
For the other cases the power $\kappa$ changes from $0.5953$ for $\zeta=1$ to $0.5020$ for $\zeta=3.9$ in accordance
with the infrared analysis. The ultraviolet properties of the solutions are slightly disturbed
compared to the cases $\alpha=0$ and $\alpha=2$. An  analysis of the ultraviolet
behaviour done similarly to the one in ref.~\cite{Fischer:2002hn} 
reveals that the $\alpha$-term in the ghost loop induces a spurious dependence
of the anomalous dimensions on the parameter $\zeta$:
\beqa
\gamma &=& \frac{-26-(\zeta-1)\alpha (2- \alpha)}{44+(\zeta-1)\alpha(2-\alpha)} \nonumber\\
\delta &=& \frac{-9}{44+(\zeta-1)\alpha (2- \alpha)}.
\eeqa
For general $\alpha$ only the transverse projector removes the spurious term in the 
ghost equation and leads to the correct one loop scaling of the equations, that is $\delta=-9/44$ 
for the ghost and $\gamma=-13/22$ for the gluon dressing function for an arbitrary number of colors and
zero flavours.

\section{Conclusion}

We have studied the infrared behaviour of the ghost and gluon propagators in
general covariant gauges. These gauges allow to interpolate via a
second gauge parameter between the linear-covariant ones of standard
Faddeev-Popov theory and the ghost-antighost symmetric gauges.
We derived the corresponding generalised Dyson--Schwinger equations
for the propagators which include the ones of linear-covariant gauges
as the limit where the second gauge parameter vanishes.  
Note that 
ghost-antighost symmetric gauges are particularly interesting as they allow an
interpretation of the antighost field being the antiparticle of the ghost which
includes also the possibility of ghost-antighost condensate. Due to the
emergence of a four-ghost interaction term in the Lagrangean for general values
of gauge parameters the Dyson--Schwinger equation of the ghost propagator
displays a rich  structure very similar to the one of the gluon equation. On
the other hand, in the gluon equation we obtain the same structure as in linear
covariant gauges apart from the fact that the gluon propagator acquires a
nontrivial longitudinal part which appears in turn in all
diagrams. The gluon
and ghost equations depend therefore on three dressing functions,
one for the ghost, one for the transverse part of the gluon propagator and one
for the longitudinal one, which are constrained, however, by
Slavnov-Taylor identities in an intricate way.

We then employed a truncation scheme for the Dyson--Schwinger equations that
uses bare vertices in place of the dressed ones. The success of this particular
truncation scheme in Landau gauge has been attributed to the
non-renormalization of the ghost-gluon vertex, that is $\widetilde Z_1=1$.  We
addressed the infrared behaviour of the ghost and gluon propagators for general
gauges by employing power law {\it ans\"atze} for the respective dressing
functions. We then have been able to evaluate the infrared behavior of the
gluon and ghost equations analytically. 

For all linear covariant gauges we find a similar result as compared to the
one  in Landau gauge: An infrared suppressed gluon propagator and an infrared
enhanced ghost. Whereas in Landau gauge there are indications that this generic
result is not changed when the vertices are dressed \cite{Lerche:2002ep}, it
remains an open question whether this is the case in linear covariant gauges in
general. Away from linear covariant gauges, that is in the general case $\alpha
\not=0$ and $\lambda \not=0$, we do not find power solutions for the dressing
functions.
However, we expect this to change with appropriate vertex dressings.
Nevertheless, it remains to be emphasized that therefore also
the occurrence of a ghost and/or gluon mass is excluded in this 
specific truncation scheme within this class of gauges.
A Dyson--Schwinger equation based investigation of the related question of a
ghost-antighost vacuum  condensate, or more generally, of an
`on-shell'-BRS-invariant dimension two condensate, needs to take into account
the generalized Slavnov--Taylor identities~(\ref{glSTI}) and (\ref{glSTIa}).
The question arises whether an infrared massless-like longitudinal part of the 
gluon propagator leads for all values of the gauge parameters to the same 
infrared enhancement of ghosts as observed in the Landau gauge. Work in this 
direction is in progress.

A special case among all gauges considered here is Landau gauge. In the limit
$\lambda=0$ the general Lagrangean (\ref{Lagrangian}) becomes independent of
the second gauge parameter $\alpha$, thus Landau gauge is also a special case
of ghost-antighost symmetric gauges. Although the Lagrangean of the theory is
independent of the gauge parameter $\alpha$, our simple truncation scheme
breaks this invariance and spurious  $\alpha$-dependent terms arise in the
ghost loop of the gluon Dyson--Schwinger equation. Examining the case
$\alpha=1$ we showed that the influence of these spurious terms is very small.
We determined solutions for the ghost and gluon dressing functions both
analytically in the infrared and numerically for finite momenta and  found
solutions identical to the ones of ref.\ \cite{Fischer:2002hn} provided the
gluon equation is projected onto its physical, transversal components. We thus
recovered the results of Landau gauge from a different direction in the two
dimensional space of gauge parameters. 

\section{Acknowledgments}

\bigskip

We are grateful to Jacques Bloch for checking our FORM code with some of his results. 
Furthermore we are indebted to Herbert Weigel for pointing out an inconsistency in the original
version of the manuscript. 
This work has been supported by the DFG under contracts Al 279/3-3, Al 279/3-4, SM 70-1/1  
and Re 856-5/1 and by the contract GRK683 (European graduate school Basel--T\"ubingen).

\newpage
\begin{appendix}

\section{Appendix A: Derivation of the Dyson--Schwinger equation for the
Ghost Propagator}

We start by transforming the Lagrangean (\ref{Lagrangian}) into a more suitable
form by partial integration, assuming the usual boundary conditions of
vanishing fields at infinity. 
In order to keep notation on a readable level we will suppress renormalization
constants in this appendix:  The derivation of the
Dyson--Schwinger equation for the ghost propagator  remains formally unchanged
by the rescaling (\ref{renorm}) and thus the appropriate renormalization
constants can be regained straightforwardly.
We obtain
\beqa
\cal{L} &=& \frac{1}{2} A_\mu^a \left( -\partial^2 \delta_{\mu \nu} + \left(1-\frac{1}{\lambda}\right)
\partial_\mu \partial_\nu \right) A_\nu^a 
- g f^{abc} \left(\partial_\mu A_\nu^a\right) A_\mu^b A_\nu^c \nonumber\\
&&+\frac{g^2}{4} f^{abe} f^{cde} A_\mu^a A_\nu^b A_\mu^c A_\nu^d + \bar{c}^a \partial^2 c^a 
+ \frac{\alpha}{2} \left(1-\frac{\alpha}{2}\right) \frac{\lambda}{2}
g^2 f^{ace} f^{bde} \bar{c}^a  \bar{c}^b c^c c^d \nonumber\\
&&+ i\left(1-\frac{\alpha}{2} \right) g f^{abc}  \bar{c}^a \partial_\mu \left(A_\mu^c c^b\right)
+ i\frac{\alpha}{2} g f^{abc} \bar{c}^a A_\mu^c \partial_\mu c^b.
\eeqa
The partition function of the theory is given by
\beq
Z[J,\sigma,\bar{\sigma}] = \int {\cal D} [A \bar{c} c] \exp\left\{-\int d^4z \, {\cal{L}} + \int d^4z \, 
\left(A^a J^a+\bar{\sigma}c+\bar{c}\sigma \right)\right\}
\eeq
with the sources $J$, $\sigma$ and $\bar{\sigma}$ of the gluon, antighost and ghost fields, respectively.
The action is given by $\cS[J,c,\bar{c}]=\int d^4z\, {\cal{L}}$.
The generating functional of connected Green's functions, $W[J,\sigma,\bar{\sigma}]$,
is defined as the logarithm of the partition function. The functional Legendre transform of $W$ is 
the effective action
\beq
\Gamma[A,\bar{c},c] = -W[J,\sigma,\bar{\sigma}] + \int d^4z \left(A^a J^a+\bar{\sigma}c+\bar{c}\sigma \right) ,
\eeq
which is the generating functional of one-particle irreducible vertex functions.
The fields and sources can be written as functional derivatives of the respective generating functionals
in the following way
\beqa
\frac{\delta W}{\delta \sigma} &=& \bar{c}, \hspace*{1cm}\frac{\delta W}{\delta \bar{\sigma}} = c, 
\hspace*{1cm}\frac{\delta W}{\delta J_\mu} = A_\mu, \nonumber\\
\frac{\delta \Gamma}{\delta c} &=& \bar{\sigma}, \hspace*{1cm}
\frac{\delta \Gamma}{\delta \bar{c}} = \bar{\sigma}, 
\hspace*{1cm}\frac{\delta \Gamma}{\delta A_\mu} = J_\mu .
\label{derivatives}
\eeqa
The sign conventions have been chosen such that
derivatives with respect to $\bar{c}$ and $\bar{\sigma}$ are left derivatives whereas
the ones with respect to ${c}$ and ${\sigma}$ are right derivatives,
\beq
\frac{\delta}{\delta\left( \bar{\sigma},\bar{c}\right)} := \mbox{left derivative} \hspace*{1cm}
\frac{\delta}{\delta\left( {\sigma},{c}\right)} := \mbox{right derivative}.
\eeq

Given that the functional integral is well-defined, the Dyson-Schwinger equation for the ghost propagator
is derived from the observation that the integral of a total derivative vanishes provided the
measure is invariant under field translations. We take the derivative 
with respect to the antighost field and obtain
\beqa
0 &=& \int {\cal D} [A \bar{c} c] \frac{\delta}{\delta \bar{c}} \exp\left\{-\int d^4z \, {\cal{L}} + \int d^4z 
\, \left(A^a J^a+\bar{\sigma}c+\bar{c}\sigma \right)\right\} \nonumber\\
&=& \int {\cal D} [A \bar{c} c] \left(-\frac{\delta S\left[A,c,\bar{c}\right]}{\delta \bar{c}} + \sigma \right)
 \exp\left\{-\int d^4z\, {\cal{L}} + \int d^4z \, 
\left(A^a J^a+\bar{\sigma}c+\bar{c}\sigma \right)\right\}\nonumber\\
&=& \left(-\frac{\delta S\left[\frac{\delta}{\delta J},\frac{\delta}{\delta \bar{\sigma}}
,\frac{\delta}{\delta \sigma}\right]}{\delta \bar{c}} + \sigma \right)Z[J,\sigma,\bar{\sigma}].
\eeqa
Now we use the relations (\ref{derivatives}) and apply a further functional derivative with
respect to the source $\sigma^b(y)$. We arrive at
\beq
0 = \left(-\frac{\delta S}{\delta \bar{c}^c(z)} \bar{c}^b(y) + \sigma^c(z) \bar{c}^b(y) + \delta(z-y)\delta_{cb}
\right)Z[J,\sigma,\bar{\sigma}]
\eeq
with explicit colour indices and space-time arguments. Setting the sources equal to zero we obtain the
ghost Dyson-Schwinger equation
\beq
\left\langle \frac{\delta S}{\delta \bar{c}^c(z)} \bar{c}^b(y) 
\right\rangle = \delta(z-y) \delta_{cb}.
\label{ghostdse}
\eeq
The derivative is easily calculated
\beqa
\frac{\delta S}{\delta \bar{c}^c(z)} &=&  \partial^2 c^c(z)
+ \frac{\alpha}{2} \left(1-\frac{\alpha}{2}\right) \frac{\lambda}{2}
g^2 f^{cde} f^{fge} \bar{c}^d(z) c^f(z) c^g(z) \nonumber\\
&&+ i\left(1-\frac{\alpha}{2} \right) g f^{cde} \partial_\mu \left(
A^e_\mu(z) c^d(z)\right)
+ i\frac{\alpha}{2} g f^{cde} A^e_\mu(z) \partial_\mu 
c^d(z).
\label{ghostder}
\eeqa
Whereas in the covariant formalism full and connected {\it three-point functions} are the same, 
the {\it four-point correlations} 
have to be decomposed into disconnected and connected parts. For the four-ghost correlation
function this results in
\beqa 
\langle \bar{c}^b(y) \bar{c}^d(z) c^f(z) c^g(z)\rangle &=& \langle \bar{c}^b(y)c^g(z)\rangle \, \langle \bar{c}^d(z) c^f(z)\rangle
 - \langle \bar{c}^b(y)c^f(z)\rangle \, \langle \bar{c}^d(z)c^g(z)\rangle \nonumber\\
&+& \langle \bar{c}^b(y) \bar{c}^d(z) c^f(z) c^g(z)\rangle_{conn.} \,.
\eeqa
Keeping in mind the Grassmann nature of the ghost and antighost fields we then obtain
\beqa
-\delta(z-y) \delta_{cb}&=&\partial^2 \langle  \bar{c}^b(y)  c^c(z) \rangle 
+ \frac{\alpha}{2} \left(1-\frac{\alpha}{2}\right) \frac{\lambda}{2}
g^2 f^{cde} f^{fge} \left\{ \langle \bar{c}^b(y) \bar{c}^d(z) c^f(z) c^g(z)\rangle \right. \nonumber\\ 
&&\left.+\left( \langle \bar{c}^b(y)c^g(z)\rangle \, \langle \bar{c}^d(z) c^f(z)\rangle
 - \langle \bar{c}^b(y)c^f(z)\rangle \, \langle \bar{c}^d(z)c^g(z)\rangle \right) \right\}\nonumber\\ 
&+& \left(1-\frac{\alpha}{2} \right) g f^{cde} \langle \bar{c}^b(y)\partial_\mu \left(
A^e_\mu(z) c^d(z)\right)\rangle
+ \frac{\alpha}{2} g f^{cde} \langle \bar{c}^b(y) A^e_\mu(z) \partial_\mu c^d(z) \rangle \,, 
\nonumber\\ \label{DSEa}
\eeqa
where all correlations are connected Green's functions.
We now use the relation
\beqa
\delta(y-x)\delta^{ab} = \frac{\delta \bar{\sigma}^b(y)}{\delta \bar{\sigma}^a(x)}
=\int d^4z \frac{\delta \bar{\sigma}^b(y)}{\delta \bar{c}^d(z)}\frac{\delta \bar{c}^d(z)}{\delta \bar{\sigma}^a(x)}
&=&\int d^4z \frac{\delta^2 \Gamma}{\delta \bar{c}^d(z) \delta c^b(y)} \frac{\delta^2 W}
{\delta \bar{\sigma}^a(x) \delta \sigma^d(z)} \nonumber\\
&=:& \int d^4z \left[D_G^{db}(z-y) \right]^{-1} D_G^{ad}(x-z) \nonumber\\
\eeqa
and multiply eq.~(\ref{DSEa}) 
with $-[D_G^{ac}(x-z)]^{-1}=\left[ \langle \bar{c}^c(z) c^a(x) \rangle \right]^{-1}$. We arrive at 
\beqa
[D_G^{ab}(x-y)]^{-1}
&=&\partial^2 \delta(x-y) \delta^{ab} \nonumber\\ 
&-& \frac{\alpha}{2} \left(1-\frac{\alpha}{2}\right) \frac{\lambda}{2}
g^2 f^{cde} f^{fge} \int d^4z \,[D_G^{ac}(x-z)]^{-1}  
\left\{ \langle \bar{c}^b(y) \bar{c}^d(z) c^f(z) c^g(z)\rangle \right.\nonumber\\
&&\hspace*{2.4cm}+\, \left.\langle \bar{c}^b(y)c^g(z)\rangle\,\langle \bar{c}^d(z) c^f(z)\rangle
 - \langle \bar{c}^b(y)c^f(z)\rangle\, \langle \bar{c}^d(z)c^g(z)\rangle \right\} \nonumber\\ 
&-& i\left(1-\frac{\alpha}{2} \right) g f^{cde} \int d^4z \, [D_G^{ac}(x-z)]^{-1}
\langle \bar{c}^b(y)\partial_\mu \left(
A^e_\mu(z) c^d(z)\right)\rangle \nonumber\\
&-& i\frac{\alpha}{2} g f^{cde} \int d^4z \, [D_G^{ac}(x-z)]^{-1}\langle \bar{c}^b(y) A^e_\mu(z) 
\partial_\mu c^d(z) \rangle. 
\nonumber\\ \label{DSEb}
\eeqa
Before we decompose the connected Green's functions into one particle irreducible ones we
have to take care of the space-time derivatives. Noting that
\beqa
\partial_\mu^z \frac{\delta^2 W}{\delta J^c_\mu(z) \sigma^d(z)} &=& -\int d^4u \partial_\mu^u\left(\delta(u-z)   
\right) \frac{\delta^2 W}{\delta J^c_\mu(u) \sigma^d(u)} \nonumber\\
&=& -\int d^4[uv] \partial_\mu^u\left(\delta(u-z)   
\right) \delta(u-v)\frac{\delta^2 W}{\delta J^c_\mu(v) \sigma^d(u)} 
\eeqa
with the abbreviation $d^4u \, d^4v =: d^4[uv]$, and 
\beqa
\frac{\delta }{\delta J^c_\mu(z)} \partial_\mu^z \frac{\delta W}{\sigma^d(z)}
&=& \int d^4[uv]\, \delta(u-z) \delta(u-v) 
\frac{\delta }{\delta J^c_\mu(v)} \partial_\mu^u \frac{\delta W}{\sigma^d(u)} \nonumber\\
&=& -\int d^4[uv]\, \partial_\mu^u\left(\delta(u-z)   
 \delta(u-v)\right)\frac{\delta^2 W}{\delta J^c_\mu(v) \sigma^d(u)} 
\eeqa
we can replace the derivative terms by the bare ghost-gluon vertex defined in appendix B.
The tadpole term can be treated in the following way:
\beqa
\int d^4z [D_G^{ac}(x-z)]^{-1} 
 f^{cde} f^{fge} \left\{\langle \bar{c}^b(y)c^g(z)\rangle\,\langle \bar{c}^d(z) c^f(z)\rangle
 - \langle \bar{c}^b(y)c^f(z)\rangle\,\langle \bar{c}^d(z)c^g(z)\rangle \right\}
 \nonumber\\
&&\hspace*{-14cm} =2 \int d^4z\, [D_G^{ac}(x-z)]^{-1} 
 f^{cde} f^{fge} \left\{\langle \bar{c}^b(y)c^g(z)\rangle\,\langle \bar{c}^d(z) c^f(z)\rangle
 \right\}
 \nonumber\\
&&\hspace*{-14cm} =2 \int d^4[zuv]\, [D_G^{ac}(x-z)]^{-1} \delta(z-u)\,\delta(u-v)\,
 f^{cde} f^{fge} \,D_G^{gb}(z-y) \,D_G^{fd}(v-u)
 \nonumber\\
&&\hspace*{-14cm} =2 \int d^4[uv]\, \delta(x-y)\,\delta(z-u)\,\delta(u-v)\, f^{bde} f^{fae}\, D_G^{fd}(v-u).
\eeqa
Plugging the expressions for the ghost-gluon loop and the one for the tadpole into eq.~(\ref{DSEb}) 
and using the expression for
the bare four-ghost vertex given in appendix B we obtain
\beqa
[D_G^{ab}(x-y)]^{-1}
&=&\partial^2 \delta(x-y) \delta^{ab} - \int d^4[uv]\, \Gamma_{4gh}^{(0) bdfa}(x,u,v,y) \, D_G^{fd}(v-u) \nonumber\\
&&+ \frac{\alpha}{2} \left(1-\frac{\alpha}{2}\right) \frac{\lambda}{2}
g^2 f^{cde} f^{fge} \times \nonumber\\
&& \hspace*{0.5cm}\int d^4[zuv] \, \delta(z-u)\, \delta(u-v) \,[D_G^{ac}(x-z)]^{-1}\:  
\langle \bar{c}^b(y) \bar{c}^d(z) c^f(u) c^g(v)\rangle \nonumber\\ 
&&- \int d^4[zuv] \, \Gamma^{(0) cde}_\mu (z,u,v) \, [D_G^{ac}(x-z)]^{-1} \, \langle \bar{c}^b(y)
A^e_\mu(v) c^d(u)\rangle \,.
\nonumber\\ \label{DSEd}
\eeqa
 
To decompose the connected Green's functions into one-particle irreducible ones we use the
relations
\beqa
\langle A^e_\mu(v) \bar{c}^b(y) c^d(u)\rangle &=&
\int d^4 [z_1z_2z_3]\, D_{\mu \nu}^{ef}(v-z_1) \,D_G^{bg}(y-z_2) \, 
\Gamma_\nu^{fhg}(z_1,z_3,z_2) \, D_G^{hd}(u-z_3)\nonumber\\ 
\\
\langle \bar{c}^b(y) \bar{c}^d(z) c^f(u) c^g(v)\rangle  &=& 
\int d^4[u_1u_2u_3u_4u_5u_6]\, \left\{ 
D_{\mu \nu}^{ek}(u_1-u_4) \, D_G^{fl}(u-u_5) \right. \nonumber\\
&&\hspace*{1cm}\times \,
\Gamma^{khl}_\nu(u_4,u_6,u_5) \, D_G^{hb}(u_6-y) \,D_G^{gi}(v-u_2) \nonumber\\
&&\hspace*{1cm}\left. \times \,
\Gamma^{eji}_\mu(u_1,u_3,u_2) \, D_G^{jd}(u_3-z) \right\}\nonumber\\
&-& \int d^4[u_1u_2u_3u_4u_5]\, \left\{
D_G^{fe}(u-u_4) \,D_G^{hb}(u_5-y) \right.\nonumber\\
&&\hspace*{1.0cm}\left. \times \,
D_G^{gi}(v-u_2) \, \Gamma_{4gh}^{jhei}(u_3,u_5,u_4,u_2) \, D_G^{jd}(u_3-z) \right\} \,, \hspace*{1.5cm}
\eeqa
which have been derived in appendix B.

Substituting these expressions into eq.~(\ref{DSEd}) we arrive at the final expression for the 
ghost Dyson-Schwinger equation in coordinate space:
\beqa
[D_G^{ab}(x-y)]^{-1}
&=&[D_G^{(0) ab}(x-y)]^{-1} \nonumber\\ 
&-&  \int d^4[uv]\, \Gamma_{4gh}^{(0) bdfa}(x,u,v,y) \, D_G^{fd}(v-u) \nonumber\\
&-& \frac{1}{2} \int d^4[zuvu_1u_2u_3u_4u_5]\, \Gamma^{(0) bdgf}_{4gh}(y,z,v,u)\, 
D_{\mu \nu}^{ek}(u_1-u_4)\, D_G^{fl}(u-u_5) \nonumber\\
&&  \hspace*{1cm} \times\, \Gamma^{kal}_\nu(u_4,x,u_5)\, D_G^{gi}(v-u_2) \,\Gamma^{eij}_\mu(u_1,u_3,u_2)
\, D_G^{jd}(u_3-z) \nonumber\\ 
&-& \frac{1}{2} \int d^4[zuvu_1u_2u_3u_4] \Gamma^{(0) bdgf}_{4gh}(y,z,v,u)\, 
D_G^{fe}(u-u_4)\nonumber\\
&& \hspace*{1cm} \times\,
D_G^{gi}(v-u_2) \,\Gamma_{4gh}^{jaei}(u_3,x,u_4,u_2) \,D_G^{jd}(u_3-z) \nonumber\\ 
&-& \int d^4[zuvz_1z_2z_3]\, \Gamma^{(0) bde}_\mu (y,u,v) \, 
D_{\mu \nu}^{ef}(v-z_1) \,\Gamma_\nu^{fha}(z_1,z_3,x)D_G^{hd}(u-z_3)
\nonumber\\ \label{DSEe}
\eeqa
where an additional minus signs arises from the interchange of the colour indices $f$ and $g$ in the
bare four-ghost vertices and from the interchange of $j$ and $i$ in the ghost-gluon vertex.

After performing a Fourier transformation we obtain the respective expression in momentum space
\beqa
[D_G(p)]^{-1}
&=&[D_G^{(0)}(p)]^{-1} \nonumber\\ 
&+& \left( -N_c \right) \frac{g^2}{(2\pi)^4} \int d^4q \: \Gamma_{4gh}^{(0)} \:  D_G(q) \nonumber\\
&+& \left( \frac{-N_c^2}{2} \right) \frac{1}{2} \frac{g^4}{(2\pi)^8} \int d^4[q_1q_2] \: \Gamma^{(0)}_{4gh}\:  
D_{\mu \nu}(p-q_1) \: D_G(q1) \nonumber\\
&&  \hspace*{1cm} \times \, \Gamma_\nu(p,q_1)\:  D_G(q_2)\: \Gamma_\mu(-p+q_1+q_2,q_2)\: D_G(p-q_1-q_2) \nonumber\\ 
&-& \left( -N_c^2 \right)  \frac{1}{2} \frac{g^4}{(2\pi)^8}\int d^4[q_1q_2] \Gamma^{(0)}_{4gh} \:
D_G(q_1) \:D_G(p-q_1-q_2) \:\Gamma_{4gh}(p,q_1,q_2)\: D_G(q_2) \nonumber\\ 
&+& \left( -N_c \right) \frac{g^2}{(2\pi)^4}\int d^4q \:\Gamma^{(0)}_\mu (p,q) \: 
D_{\mu \nu}(p-q) \:\Gamma_\nu(q,p)D_G(q)
\nonumber\\ \label{DSE-ghost}
\eeqa
where the colour traces have been carried out and the reduced vertices defined in 
appendix B have been used.

\section{Appendix B: Definitions and decompositions}

\begin{figure}
\centerline{
\epsfig{file=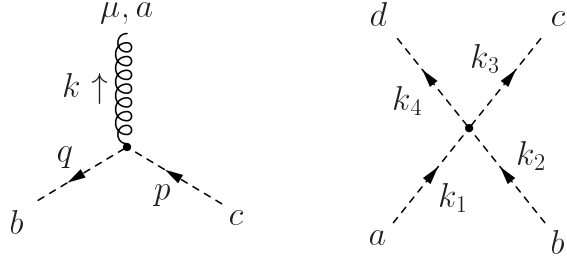,width=7.5cm}
}
\caption{\label{TL-vertices}
Momentum routing for the tree level ghost-gluon and four-ghost vertices.}
\end{figure}

{\bf Ghost and gluon propagators:}

The full ghost and gluon propagators in coordinate space are defined to be
\beqa
\langle c^a(x) \bar{c}^b(y) \rangle &=& \frac{\delta W}{\delta \sigma^b(y) \delta \bar{\sigma}^a(x)}
=D_G^{ab}(x-y) \\
\langle A_\mu^a(x) A_\nu^b(y) \rangle &=& \frac{\delta W}{\delta J_\nu^b(y) \delta J_\nu^a(x)}
=D_{\mu \nu}^{ab}(x-y). \\
\eeqa
The bare propagators in coordinate space can be easily derived from the quadratic part of the action,
\beq
S_{quad}= \int d^4x^\prime \left\{ \frac{1}{2} A_\mu^a \left( -\partial^2 \delta_{\mu \nu} 
+ \left(1-\frac{1}{\lambda}\right)
\partial_\mu \partial_\nu \right) A_\nu^a+ \bar{c}^a \partial^2 c^a \right\}
\eeq
and are given by
\beqa
\left[D_G^{(0) ab}(x-y)\right]^{-1} &=& \frac{\delta^2 S_{{quad}}}{\delta \bar{c}^a(x) c^b(y)} 
 \hspace*{3mm}=\delta^{ab} \partial^2  \delta(x-y) \,, \\
\left[D_{\mu \nu}^{(0) ab}(x-y)\right]^{-1} &=&\frac{\delta^2 S_{{quad}}}{\delta A_\mu^a(x) A_\nu^b(y)} 
= \delta^{ab}\left(- \partial^2 \delta_{\mu \nu} + \left(1-\frac{1}{\lambda}\right)\partial_\mu 
\partial_\nu \right) \, \delta(x-y)\,, \hspace*{1cm} 
\eeqa
with the gauge parameter $\lambda$.
After Fourier transformation one obtains the corresponding expressions in momentum space:
\beqa
\left[D_G^{(0) ab}(p)\right]^{-1} &=& -\delta^{ab} p^2 \\
\left[D_{\mu \nu}^{(0) ab}(p)\right]^{-1} &=& \delta^{ab}\left(\delta_{\mu \nu} 
     - \left(1-\frac{1}{\lambda}\right) \frac{p_\mu p_\nu}{p^2}\right)p^2.
\eeqa

{\bf Ghost-gluon vertex:}

From the ghost gluon part of the action
\beq
S_{ghgl} =  \int d^4x^\prime \left\{ -i\left(1-\frac{\alpha}{2} \right) g f^{abc} 
\left(\partial^\mu \bar{c}\right) A^{c}_\mu c^b
+ i\frac{\alpha}{2} g f^{abc} \bar{c^a} A_\mu^c \partial^\mu c^b \right\}
\eeq
the tree level ghost gluon vertex $\Gamma_\mu^{abc}$ is easily derived:
\beqa
\Gamma_\mu^{(0) abc}(x,y,z) &=& \frac{\delta^3 S_{ghgl}}{\delta A_\mu^a(x) \delta \bar{c}^b(y) \delta c^c(z)} \nonumber\\
&=& -g f^{abc} \left[ i\left(1-\frac{\alpha}{2}\right) \left(\partial_\mu^z \delta^4(z-y) \right) \delta^4(z-x)
+ i\frac{\alpha}{2} \partial_\mu^z\left(\delta^4(z-y) \delta^4(z-x) \right)  \right]. \nonumber\\
\eeqa
Using the momentum conventions of Fig.~\ref{TL-vertices} the Fourier transformed 
bare ghost-gluon vertex reads 
\beqa
\Gamma_\mu^{(0) abc}(k,p,q) &=& \int d^4 [x y z] \, \Gamma_\mu^{abc}(x,y,z) \, e^{i(k\cdot x+q\cdot y-p\cdot z)} \nonumber\\
&=& g f^{abc}\, (2 \pi)^4 \,\delta^4(k+q-p) \,\left[ \left(1-\frac{\alpha}{2}\right)q_\mu 
+ \frac{\alpha}{2} p_\mu \right] \,,
\eeqa
where the abbreviation $d^4x\, d^4y\, d^4z =: d^4[xyz]$ has been introduced.
Note the symmetry of the vertex in the ghost momenta $p_\mu$ and $q_\mu$ if $\alpha=1$.
For convenience we define a reduced vertex function $\Gamma_\mu^{(0)}(p,q)$ by
\beqa
\Gamma_\mu^{(0) abc}(k,p,q) &=& g f^{abc} (2 \pi)^4 \delta^4(k+q-p) \Gamma_\mu^{(0)}(p,q) \nonumber\\
\Gamma_\mu^{(0)}(p,q) &=& \left[ \left(1-\frac{\alpha}{2}\right)q_\mu + \frac{\alpha}{2} p_\mu \right].
\eeqa
The full one particle irreducible ghost gluon vertex in coordinate space is given by
\beq
\Gamma_\mu^{abc}(x,y,z) = 
\frac{\delta \Gamma}{
\delta J^a_\mu(x) \delta \bar{c}^b(y) \delta {c}^c(z)}.
\eeq

\pagebreak
{\bf Four-ghost vertex:}

The four-ghost vertex $\Gamma_{4g}^{abcd}$ is derived from the four ghost part of the action
\beq
S_{4gh} =  \int d^4x^\prime \left\{\frac{\alpha}{2} \left(1-\frac{\alpha}{2}\right) \frac{\lambda}{2}
g^2 f^{ace} f^{bde} \bar{c}^a  \bar{c}^b c^c c^d \right\}
\eeq
which leads to
\beqa
\Gamma_{4g}^{(0) abcd}(x,y,z,w) &=& \frac{\delta^4 S_{4gh}}{\delta \bar{c}^a(x) \delta \bar{c}^b(y) 
\delta c^c(z)\delta c^d(w)} \nonumber\\
&=&  \frac{\alpha}{2} \left(1-\frac{\alpha}{2}\right) \lambda 
g^2 f^{abe} f^{cde} \delta^4(x-y) \delta^4(y-z) \delta^4(z-w).
\eeqa
Again using the momentum conventions of figure (\ref{TL-vertices}) one obtains for the Fourier-transformed 
bare four-ghost vertex
\beqa
\Gamma_{4g}^{(0) abcd}(k_1,k_2,k_3,k_4) &=& \frac{\delta^4 S_{4gh}}{\delta \bar{c}^a(x) \delta \bar{c}^b(y) 
\delta c^c(z)\delta c^d(w)} \nonumber\\
&=&  \frac{\alpha}{2} \left(1-\frac{\alpha}{2}\right) \lambda 
g^2 f^{abe} f^{cde} (2\pi)^4 \delta^4(k_1+k_2-k_3-k_4).
\eeqa
We define a reduced vertex function $\Gamma_{4g}^{(0)}$ by
\beqa
\Gamma_{4g}^{(0) abcd}(k_1,k_2,k_3,k_4) &=& 
g^2 f^{abe} f^{cde} (2\pi)^4 \delta^4(k_1+k_2-k_3-k_4) \Gamma_{4g}^{(0)}\nonumber\\
\Gamma_{4g}^{(0)} &=& \frac{\alpha}{2} \left(1-\frac{\alpha}{2}\right) \lambda. \label{barefourghostvertex} 
\eeqa 
The full four-ghost vertex in coordinate space is formally given by
\beq
\Gamma^{abcd}(x,y,z) = 
\frac{\delta \Gamma}{\delta \bar{c}^a(x) \delta \bar{c}^b(y) \delta {c}^c(z) \delta {c}^d(y)}.
\eeq

{\bf Decomposition of connected ghost-gluon Green's function:}

With the help of the matrix relation
\beq
\frac{\delta \chi^{-1}}{\delta \phi}  = - \chi^{-1} \frac{\delta \chi}{\delta \phi} \chi^{-1},
\eeq
and the identity
\beqa
\delta(y-x)\delta^{ab}  
=\int d^4z \frac{\delta \bar{\sigma}^b(y)}{\delta \bar{c}^d(z)}\frac{\delta \bar{c}^d(z)}{\delta \bar{\sigma}^a(x)}
&=&\int d^4z \frac{\delta^2 \Gamma}{\delta \bar{c}^d(z) \delta c^b(y)} \frac{\delta^2 W}
{\delta \bar{\sigma}^a(x) \delta \sigma^d(z)} \,, 
\eeqa
we decompose the connected ghost-gluon correlation function, 
$\langle A^a_\mu(x) \bar{c}^b(y) c^c(z) \rangle$, in the following way:
\beqa
\langle A^a_\mu(x) \bar{c}^b(y) c^c(z) \rangle &=& 
\frac{\delta^3 W}{\delta J^a_\mu(x) \delta \bar{\sigma}^b(y) \delta {\sigma}^c(z)} \nonumber\\
&&\hspace*{-3cm}= \frac{\delta}{\delta J^a_\mu(x)} \left[ 
\frac{\delta^2 \Gamma}{\delta \bar{c}^b(y) \delta {c}^c(z)}\right]^{-1} \nonumber\\ 
&&\hspace*{-3cm}= 
\int d^4u_1 \frac{\delta A^d_\nu(u_1)}{\delta J^a_\mu(x)}\frac{\delta}{\delta A^d_\nu(u_1)}
 \left[ \frac{\delta^2 \Gamma}{\delta \bar{c}^b(y) \delta {c}^c(z)}\right]^{-1} \nonumber\\
&&\hspace*{-3cm}= \int d^4[u_1u_2u_3] \frac{\delta^2 W}{\delta J^a_\mu(x) \delta J^d_\nu(u_1)} \:
\frac{\delta^2 W}{\delta \bar{\sigma}^b(y) \delta {\sigma}^e(u_2)} \: \frac{\delta^3 \Gamma}{
\delta A^d_\nu(u_1) \delta \bar{c}^e(u_2) \delta {c}^f(u_3)} \:
\frac{\delta^2 W}{\delta \bar{\sigma}^f(u_3) \delta {\sigma}^c(z)}\nonumber\\ 
&&\hspace*{-3cm}=\int d^4 [u_1u_2u_3] D_{\mu \nu}^{ad}(x-u_1) D_G^{eb}(u_2-y) 
\Gamma^{def}_\nu(u_1,u_2,u_3)D_G^{cf}(u_3-z).
 \label{ghgldecomp}
\eeqa
Here we used the abbreviation $d^4[u_1u_2u_3]:=d^4u_1\, d^4 u_2 \, d^4u_3$ and the definitions of the
gluon propagator $D_{\mu \nu}$, the ghost propagator $D_G$ and the ghost-gluon vertex $\Gamma_\nu$
given in previous subsections.

{\bf Decomposition of connected four ghost Green's function:}

Furthermore we need the decomposition of the four-ghost correlation function into
one-particle irreducible parts. We start at a stage where the sources are still present and set them
to zero at the end of the derivation.
We first give the decomposition of the connected ghost-antighost-ghost
three-point function
\beqa
\langle \bar{c}^b(y) {c}^c(z) \bar{c}^d(w) \rangle &=& 
\frac{\delta^3 W}{\delta \bar{\sigma}^b(y) \delta \sigma^c(z) \delta \bar{\sigma}^d(w)} \nonumber\\
&&\hspace*{-3cm}= \frac{\delta}{\delta \bar{\sigma}^b(y)} \left[ \frac{\delta^2 \Gamma}
{\delta \sigma^c(z) \delta \bar{\sigma}^d(w)}\right]^{-1} \nonumber\\
&&\hspace*{-3cm}=\int d^4u_1 \frac{\delta A^e_\nu(u_1)}{\delta \bar{\sigma}^b(y)}\frac{\delta}{\delta A^e_\nu(u_1)}
 \left[ \frac{\delta^2 \Gamma}{\delta \sigma^c(z) \delta \bar{\sigma}^d(w)}\right]^{-1} \nonumber\\
&&\hspace*{-3cm}= \int d^4[u_1u_2u_3] \frac{\delta^2 W}{\delta \bar{\sigma}^b(y) \delta J^e_\nu(u_1)} \:
\frac{\delta^2 W}{\delta \sigma^c(z) \delta \bar{\sigma}^f(u_2)} \: \frac{\delta^3 \Gamma}{
\delta A^e_\nu(u_1) \delta {c}^f(u_2) \delta \bar{c}^g(u_3)} \:
\frac{\delta^2 W}{\delta \sigma^g(u_3) \delta \bar{\sigma}^d(w)} \,. \nonumber\\
\eeqa
Then we decompose the connected four-ghost Green's function:
\beqa
\langle {c}^a(x) \bar{c}^b(y) {c}^c(z) \bar{c}^d(w) \rangle &=& \frac{\delta^4 W}
{\delta \sigma^a(x) \delta \bar{\sigma}^b(y) \delta \sigma^c(z) \delta \bar{\sigma}^d(w)} \nonumber\\
&=& \frac{\delta}{\delta \sigma^a(x)}
\int d^4[u_1u_2u_3] \frac{\delta^2 W}{\delta \bar{\sigma}^b(y) \delta J^e_\mu(u_1)} \:
\frac{\delta^2 W}{\delta \sigma^c(z) \delta \bar{\sigma}^f(u_2)} \nonumber\\
&& \hspace*{1.5cm}\times\frac{\delta^3 \Gamma}{
\delta A^e_\mu(u_1) \delta {c}^f(u_2) \delta \bar{c}^g(u_3)} \:
\frac{\delta^2 W}{\delta \sigma^g(u_3) \delta \bar{\sigma}^d(w)}.\hspace*{1cm}
\eeqa
Carrying out the remaining derivative gives four terms. The two terms where the derivative acts on the second and
on the last propagator vanish, because the term $\frac{\delta W}{\delta \bar{\sigma}^b(y) \delta J^e_\nu(u_1)}$
vanishes when the sources are set to zero. The contribution where the derivative acts on the first propagator
can be treated using eq.~(\ref{ghgldecomp}). In the expression  with the derivative acting on the vertex we
use
\beqa
-\frac{\delta^2 W}{\delta \bar{\sigma}^b(y) \delta J^e_\mu(u_1)} 
\frac{\delta^4 \Gamma}{\delta \sigma^a(x) \delta A^e_\mu(u_1) \delta {c}^f(u_2) \delta \bar{c}^g(u_3)}
&=& \frac{\delta^4 \Gamma}{\delta \sigma^a(x) \delta \bar{\sigma}^b(y) \delta {c}^f(u_2) \delta \bar{c}^g(u_3)} 
\nonumber\\
&&\hspace*{-7cm}= \int d^4u_4\frac{\delta^2 W}{\delta \sigma^a(x) \bar{\sigma}^e(u_4)}\:\:
\frac{\delta^4 \Gamma}{\delta c^e(u_4) \delta \bar{\sigma}^b(y) \delta {c}^f(u_2) \delta \bar{c}^g(u_3)}
\nonumber\\
&&\hspace*{-7cm}= \int d^4[u_4u_5]\frac{\delta^2 W}{\delta \sigma^a(x) \bar{\sigma}^e(u_4)}\:\: 
\frac{\delta^2 W}{\delta \bar{\sigma}^b(y) \delta{\sigma}^h(u_5)}\:\:
\frac{\delta^4 \Gamma}{\delta c^e(u_4) \delta \bar{c}^h(u_5) \delta {c}^f(u_2) \delta \bar{c}^g(u_3)}
\nonumber\\
&&\hspace*{-7cm}= -\int d^4[u_4u_5]\frac{\delta^2 W}{\delta \sigma^a(x) \bar{\sigma}^e(u_4)} \:\:
\frac{\delta^2 W}{\delta{\sigma}^h(u_5) \delta \bar{\sigma}^b(y)}\:\:
\frac{\delta^4 \Gamma}{ \delta \bar{c}^g(u_3)\delta \bar{c}^h(u_5)\delta c^e(u_4) \delta {c}^f(u_2)}.
\nonumber\\
\eeqa
Collecting all this together we arrive at
\beqa
\langle {c}^a(x) \bar{c}^b(y) {c}^c(z) \bar{c}^d(w) \rangle &=&
\int d^4[u_1u_2u_3u_4u_5u_6] \left\{ 
\frac{\delta^2 W}{\delta J^e_\mu(u_1)\delta J^f_\nu(u_4)} \:\:
\frac{\delta^2 W}{\delta \sigma^a(x) \delta \bar{\sigma}^g(u_5)} \right. \nonumber\\
&&\hspace*{1cm} \times
\frac{\delta^3 \Gamma}{\delta A^f_\nu(u_4) \delta {c}^g(u_5) \delta \bar{c}^h(u_6)}\:\:
\frac{\delta^2 W}{\delta \sigma^h(u_6) \delta \bar{\sigma}^b(y)}\:\:
\frac{\delta^2 W}{\delta \sigma^c(z) \delta \bar{\sigma}^i(u_2)} \nonumber\\
&&\hspace*{1cm} \left. \times
\frac{\delta^3 \Gamma}{\delta A^e_\mu(u_1) \delta {c}^i(u_2) \delta \bar{c}^j(u_3)}\:\:
\frac{\delta^2 W}{\delta \sigma^j(u_3) \delta \bar{\sigma}^d(w)}\right\}\nonumber\\
&-& \int d^4[u_1u_2u_3u_4u_5] \left\{
\frac{\delta^2 W}{\delta \sigma^a(x) \bar{\sigma}^e(u_4)} \:\:
\frac{\delta^2 W}{\delta{\sigma}^h(u_5) \delta \bar{\sigma}^b(y)} \right.\nonumber\\
&&\hspace*{1cm}\times
\frac{\delta^2 W}{\delta \sigma^c(z) \delta \bar{\sigma}^f(u_2)}\:\:
\frac{\delta^4 \Gamma}{ \delta \bar{c}^g(u_3)\delta \bar{c}^h(u_5)\delta c^e(u_4) \delta {c}^f(u_2)} 
\nonumber\\
&&\hspace*{1cm}\times\left. \frac{\delta^2 W}{\delta \sigma^g(u_3) \delta \bar{\sigma}^d(w)}\right\}.
\eeqa
Interchanging some Grassmann fields in the correlations and using the definitions for the propagators and
vertices given in the previous subsections we arrive at
\beqa
\langle \bar{c}^b(y) \bar{c}^d(w) {c}^a(x) {c}^c(z)  \rangle &=& 
\int d^4[u_1u_2u_3u_4u_5u_6] \left\{ 
D_{\mu \nu}^{ef}(u_1-u_4)  \, D_G^{ag}(x-u_5) \right. \nonumber\\
&&\hspace*{1cm}\times\Gamma^{fhg}_\nu(u_4,u_6,u_5) D_G^{hb}(u_6-y) \, D_G^{ci}(z-u_2) \nonumber\\
&&\hspace*{1cm}\left.\times \Gamma^{eji}_\mu(u_1,u_3,u_2) \, D_G^{jd}(u_3-w) \right\}\nonumber\\
&+& \int d^4[u_1u_2u_3u_4u_5] \left\{
D_G^{ae}(x-u_4) \, D_G^{hb}(u_5-y) \right.\nonumber\\
&&\hspace*{1cm}\left. \times
D_G^{cf}(z-u_2) \, \Gamma_{4gh}^{hgef}(u_5,u_3,u_4,u_2) \, D_G^{gd}(u_3-w) \right\}, \hspace*{1cm}
\eeqa
which is the decomposition of the four-ghost correlation used in appendix A.

\section{Appendix C: Tensor integrals}

The explicit expression for the scalar bubble integral $I$, defined in eq.~(\ref{sc-bubble}), 
can be easily evaluated in Euclidean space-time
using the Feynman-parameterisation. 
With the squared momenta $x=p^2$, $y=q^2$ and $z=(p-q)^2$ the result is given by
\beqa
I(a,b,p) &:=& \int d^4q \frac{1}{y^a z^b} \label{sc-bubble}\\ 
&=& \pi^2 \,x^{2-a-b}\, \frac{\Gamma(2-a)\,\Gamma(2-b)\,\Gamma(a+b-2)}
{\Gamma(a)\,\Gamma(b)\,\Gamma(4-a-b)}\label{sc} \,.
\eeqa
The corresponding tensor integrals can be reduced to scalar integrals by extracting combinations of 
momenta $p_\mu$ and the symmetric tensor $\delta_{\mu \nu}$ according to the symmetry 
properties of the integrand:
\beqa
J_\mu(a,b,p)&:=&\int d^4q \frac{q_\mu}{y^a z^b}\hspace*{0.7cm} =J_1(a,b,p) \,\, p_\mu  \,,\\
K_{\mu \nu}(a,b,p)&:=&\int d^4q \frac{q_\mu q_\nu}{y^a z^b}\hspace*{0.6cm}\,=K_1(a,b,p)\,\,p_\mu p_\nu   
+ K_2(a,b,p) \,\,x \, \delta_{\mu \nu} \,, \\
L_{\mu\nu\rho}(a,b,p)&:=&\int d^4q \frac{q_\mu q_\nu q_\rho}{y^a z^b}\hspace*{0.3cm}\,
=L_1(a,b,p)\,\,p_\mu p_\nu p_\rho  \nonumber\\
&&\hspace*{2.7cm}\, +\,L_2(a,b,p)\,\, x \,\left(p_\mu \,\delta_{\nu \rho}+p_\nu \,\delta_{\rho \mu} 
+ p_\rho \,\delta_{\mu \nu} \right) \,,\\
M_{\mu\nu\rho\sigma}(a,b,p)&:=&\int d^4q \frac{q_\mu q_\nu q_\rho q_\sigma}{y^a z^b}=
M_1(a,b,p)\,\,p_\mu p_\nu p_\rho p_\sigma   \nonumber\\
&&\hspace*{2.8cm}+ \,M_2(a,b,p)\,\,x\,\left(\delta_{\mu \nu} \,p_\rho p_\sigma + \delta_{\mu \rho} \,p_\nu p_\sigma +
\delta_{\mu \sigma} \,p_\rho p_\mu  +\right. \nonumber\\
&&\hspace*{5.6cm}\left. \,\delta_{\nu \rho} \,p_\mu p_\sigma 
+\delta_{\nu \sigma}\, p_\rho p_\mu + \delta_{\rho \sigma} \,p_\mu p_\nu \right) \nonumber\\
&&\hspace*{2.8cm}+\,M_3(a,b,p)\,\,
x^2\, \left(\delta_{\mu \nu}\,\delta_{\rho \sigma}+\delta_{\mu \rho}\,\delta_{\nu \sigma}+
\delta_{\mu \sigma}\,\delta_{\rho \nu} \right)\,. \hspace*{1cm}
\eeqa
The scalar integrals in these expressions are calculated by contracting them with appropriate tensors,
writing all scalar products in terms of squared momenta $x,y$ and $z$
and applying eq.~(\ref{sc}). One arrives at 
\beqa
J_1&=& \pi^2 \frac{\Gamma(3-a)\,\Gamma(2-b)\,\Gamma(a+b-2)}
{\Gamma(a)\,\Gamma(b)\,\Gamma(5-a-b)}\, x^{2-a-b}       \,,  \\
K_1&=& \pi^2 \frac{\Gamma(4-a)\,\Gamma(2-b)\,\Gamma(a+b-2)}
{\Gamma(a)\,\Gamma(b)\,\Gamma(6-a-b)}\, x^{2-a-b}  \,, \\
K_2&=& \pi^2 \frac{\Gamma(3-a)\,\Gamma(3-b)\,\Gamma(a+b-2)}
{\Gamma(a)\,\Gamma(b)\,\Gamma(6-a-b)} \,\frac{1}{2(-3+a+b)}\, x^{2-a-b} \,,\\
L_1&=& \pi^2 \frac{\Gamma(5-a)\,\Gamma(2-b)\,\Gamma(a+b-2)}
{\Gamma(a)\,\Gamma(b)\,\Gamma(7-a-b)} \,x^{2-a-b}\,,\\
L_2&=& \pi^2 \frac{\Gamma(4-a)\,\Gamma(3-b)\,\Gamma(a+b-2)}
{\Gamma(a)\,\Gamma(b)\,\Gamma(7-a-b)}\,\frac{1}{2(-3+a+b)} \,x^{2-a-b}\,,\\
M_1&=& \pi^2 \frac{\Gamma(6-a)\,\Gamma(2-b)\,\Gamma(a+b-2)}
{\Gamma(a)\,\Gamma(b)\,\Gamma(8-a-b)} \,x^{2-a-b}\,,\\
M_2&=& \pi^2 \frac{\Gamma(5-a)\,\Gamma(3-b)\,\Gamma(a+b-2)}
{\Gamma(a)\,\Gamma(b)\,\Gamma(8-a-b)}\,\frac{1}{2(-3+a+b)} \, x^{2-a-b}\,,\\
M_3&=& \pi^2 \frac{\Gamma(4-a)\,\Gamma(4-b)\,\Gamma(a+b-2)}
{\Gamma(a)\,\Gamma(b)\,\Gamma(8-a-b)}\,\frac{1}{4(-3+a+b)(-4+a+b)}\, x^{2-a-b} \,.\hspace*{1cm}
\eeqa

\section{Appendix D: Expressions for some diagrams in bare vertex approximation}

In this appendix we give explicitly the expressions for some diagrams needed for our investigation in
the main body of the paper. All algebraic manipulations have been done using the program FORM 
\cite{Form}. Our {\it ans\"atze} for the small momentum behaviour of the
ghost dressing function $G$, the transversal gluon dressing function $Z$ and the longitudinal 
gluon dressing function $L$ are the power laws 
\beq
G(x) = Bx^\beta, \:\:\:\:
Z(x) = Ax^\sigma, \:\:\:\:
L(x) = Cx^\rho,
\label{ansatz}
\eeq 
where we have used the abbreviation $x=p^2$.

\begin{figure}
\centerline{
\epsfig{file=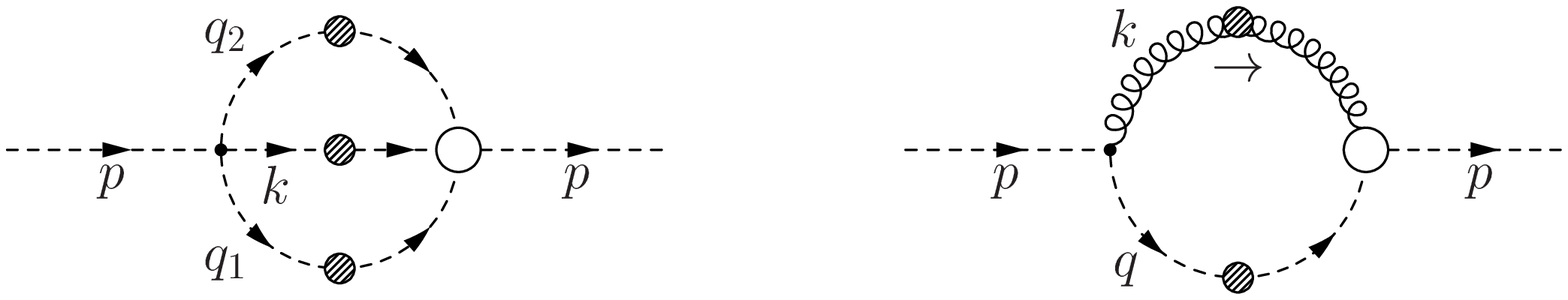,width=14cm}
}
\caption{ \sf \sf \label{ghostsunset}
Momentum routing for the sunset and for the dressing diagram in the ghost equation.}
\end{figure}

We first evaluate the sunset diagram in the ghost equation given diagrammatically in Fig.~\ref{ghostsunset}.
With the bare four-ghost vertex given in eq.~(\ref{barefourghostvertex}) and the abbreviations for the
squared momenta
$x=p^2$, $y_1=(q_1)^2$, $y_2=(q_2)^2$, $z_1=(p-q_1)^2$ and $z_2=(p-q_1-q_2)^2$ the sunset
diagram reads
\beq
U^{sun} = \frac{N_c^2 \,g^4 \,\tilde{Z}_4}{2 \,(2\pi)^8} \left(\frac{\alpha}{2}\left(1-\frac{\alpha}{2}\right)\lambda\right)^2
\int d^4q_1 \, \frac{B \,(y_1)^\beta}{x \,y_1} \,\int d^4q_2 \,\frac{B^2 \,(y_2)^\beta \,(z_2)^\beta}{y_2 \,z_2} .
\eeq
The factor $1/x$ in the first integral stems from the left hand side of the ghost equation. 
We now integrate the
inner loop with the help of formula (\ref{sc}) and obtain
\beq
U^{sun} = \frac{N_c^2\, g^4 \,\tilde{Z}_4 \,B^3}{512 \,\pi^6} \left(\frac{\alpha}{2}\left(1-\frac{\alpha}{2}\right)\lambda\right)^2
\frac{\Gamma^2(1+\beta)\,\Gamma(-2\beta)}{\Gamma^2(1-\beta)\,\Gamma(2+2\beta)}\,
\int d^4q_1  \,\frac{(y_1)^\beta}{x\, y_1}\, (z_1)^{2\beta} \,, 
\eeq
where $z_1$ is the total squared momentum flowing through the integrated loop. The second integration is done
in the same way. We arrive at
\beqa
U^{sun} &=& x^{3\beta} \,\frac{N_c^2 \,g^4 \,\tilde{Z}_4\, B^3}{512 \,\pi^4} \left(\frac{\alpha}{2}\left(1-\frac{\alpha}{2}\right)\lambda\right)^2
\frac{\Gamma^3(1+\beta)\,\Gamma(-3\beta-1)}{\Gamma^3(1-\beta)\,\Gamma(3+3\beta)}\nonumber\\
&:=& x^{3\beta} (U^\prime)^{sun} \,.
\eeqa
As each integration step eats up the two squared momenta in the denominators of the integral kernels only 
powers of $x$ to the anomalous dimensions of the dressing functions in the loop (here $3\beta$ from
three ghost propagators) survive. This mechanism works in the same way for all diagrams and explains 
the pattern in the eqs.~(\ref{gheq1}), (\ref{gl_trans1}) and (\ref{gl_long1}) in the main body of the 
paper.

Next we evaluate the two contributions in the gluon equation needed for the argument below eq.~(\ref{gl_long2}). 
The explicit expressions for the kernels of two-loop gluon diagrams are rather lengthy but the calculation
is done along the same lines as in the ghost sunset diagram above. Therefore we just give
the final results:
\beqa
V^{squint}_{TTTT} &=& x^{4\sigma} \, \frac{-27 \, g^4 \, N_c^2 \, Z_4 \, A^4}{4096 \, \pi^4}\,  \frac{\Gamma(-1-4\sigma)\:
\Gamma(1/2-\sigma)\:
\Gamma(3\sigma)\: \Gamma^2(1+\sigma)}{\Gamma(4-3\sigma)\:\Gamma^2(2-\sigma)\:\Gamma(3/2-\sigma)\:\Gamma(4+4\sigma)}
\times \nonumber\\
&& 2^{-4 \sigma}\:(-1+3\sigma)\:(10+\sigma-66\sigma^2+63\sigma^3)\:(5+43\sigma+47\sigma^2) \nonumber\\
&:=& x^{4\sigma}\, (V^\prime)^{squint}_{TTTT} \,, \\
W^{sun}_{LLL} &=&x^{3\delta}\, \frac{g^4 \, N_c^2 \, Z_4\,  C^3}{1536 \, \pi^4} \, \frac{1}{(1+3\delta)} 
\frac{\Gamma^3(1+\delta)\:\Gamma(1-3\delta)}{\Gamma^3(2-\delta)\: 
\Gamma(3+3\delta)} \:\: \lambda^3   \nonumber\\
&:=& x^{3\delta} \, (W^\prime)^{sun}_{LLL} \,.
\eeqa 

Finally we calculate that part in the dressing diagram of the ghost equation which contains the
longitudinal part of the gluon propagator for the special case $\alpha=0,2$. 
These are the linear covariant gauges where $L(x)=1$ by virtue of the Slavnov-Taylor identity.
Replacing dressed vertices with bare ones, however, violates this identity. We therefore start with the 
general expression, $L(x) = C x^\rho$ and investigate whether the limit $\rho \rightarrow 0$ can be performed
consistently. With the momentum assignments $x=p^2$, $y=q^2$ and $z=k^2=(p-q)^2$ the 
longitudinal part of the diagram is given by
\beqa
U^{dress}_L &=& -\frac{N_c \, g^2 \,\tilde{Z}_1 \,\lambda }{(2\pi)^4} \int d^4q\, q_\mu 
 \frac{k_\mu k_\nu}{z^2}p_\nu  \frac{B\, y^\beta\, C \,z^\rho}{x\,y} \\
 &=& -\frac{N_c \, g^2 \,\tilde{Z}_1 \,\lambda \,B\,C }{(2\pi)^4} \int d^4q
 \left\{ p_\mu \frac{q_\mu}{y^{1-\beta} z^{2-\rho}} - \frac{p_\mu p_\nu}{x}
\frac{q_\nu q_\mu}{y^{1-\beta}z^{2-\rho}} - \frac{1}{y^{-\beta} z^{2-\rho}} +
\frac{p_\mu}{x}\frac{q_\mu}{y^{-\beta} z^{2-\rho}}\right\} \nonumber\\
\eeqa
where again the extra factor $1/x$ stems from the left hand side of the ghost DSE. At this stage of the calculation
it is not clear whether there are infrared singularities in the limit $\rho \rightarrow 0$. 
We employ the tensor integrals given in appendix C, use
$x\Gamma(x) = \Gamma(1+x)$ and obtain
\beqa
U^{dress}_L &=& -\frac{N_c \, g^2 \,\tilde{Z}_1 \,\lambda \,B\, C }{16 \pi^2} x^{\beta+\rho} 
\frac{\Gamma(2+\beta) \Gamma(\rho) \Gamma(-\beta-\rho)}
{\Gamma(-\beta) \Gamma(2-\rho) \Gamma(2+\rho+\beta)} \frac{\rho^2 + \rho/2}{\beta (2+\beta+\rho)} \,.
\eeqa
In the limit $\rho \rightarrow 0$ this expression is infrared-finite, as 
$\lim_{\rho \rightarrow 0}\Gamma(\rho)\rho =1$. We then obtain
\beqa
U^{dress}_L &=& -\frac{N_c \, g^2 \,\tilde{Z}_1 \,\lambda \,B\, C  }{16 \pi^2} x^{\beta} 
\frac{1}{2\beta (2+\beta)} \,.
\eeqa
\end{appendix}


\newpage

\begin{thebibliography}{99}


\bibitem{Zwanziger:2003cf}
D.~Zwanziger,
arXiv:hep-ph/0303028.

\bibitem{Roberts:2000aa}
P.~Maris and C.~D.~Roberts,
arXiv:nucl-th/0301049.
\newline
C.~D.~Roberts and S.~M.~Schmidt,
Prog.\ Part.\ Nucl.\ Phys.\ {\bf 45}, S1 (2000) 
[arXiv:nucl-th/0005064].

\bibitem{Alkofer:2001wg}
R.~Alkofer and L.~von Smekal,
Phys.\ Rept.\  {\bf 353}, 281 (2001) 
[arXiv:hep-ph/0007355].

\bibitem{vonSmekal:1997is}
L.~von Smekal, R.~Alkofer and A.~Hauck,
Phys.\ Rev.\ Lett.\  {\bf 79}, 3591 (1997) \newline
[arXiv:hep-ph/9705242];
Annals Phys.\  {\bf 267}, 1 (1998)
[arXiv:hep-ph/9707327].

\bibitem{Atkinson:1998tu}
D.~Atkinson and J.~C.~Bloch,
Phys.\ Rev.\ D {\bf 58}, 094036 (1998)
[arXiv:hep-ph/9712459]; \newline
Mod.\ Phys.\ Lett.\ {\bf A13}, 1055 (1998)
[arXiv:hep-ph/9802239].

\bibitem{Bloch:2003}
J.~C.~Bloch, arXiv:hep-ph/0303125.

\bibitem{Zwanziger:2001kw}
D.~Zwanziger,
Phys.\ Rev.\ D {\bf 65}, 094039 (2002) [arXiv:hep-th/0109224].

\bibitem{Lerche:2002ep}
C.~Lerche and L.~von Smekal,
Phys.\ Rev.\ D {\bf 65}, 125006 (2002) 
[arXiv:hep-ph/0202194].

\bibitem{Fischer:2002hn}
C.~S.~Fischer  and R.~Alkofer,
Phys. Lett. {\bf B536}, 177 (2002) 
[arXiv:hep-ph/0202202]; 

C.~S.~Fischer, R.~Alkofer and H.~Reinhardt,
Phys.\ Rev.\ D {\bf 65}, 094008 (2002) 
[arXiv:hep-ph/0202195]; 

R.~Alkofer, C.~S.~Fischer and L.~von Smekal,
arXiv:hep-ph/0301107;
arXiv:nucl-th/0301048.

\bibitem{Bonnet:2000kw}
F.~D.~Bonnet, P.~O.~Bowman, D.~B.~Leinweber and A.~G.~Williams,
Phys.\ Rev.\ D {\bf 62}, 051501 (2000)
[arXiv:hep-lat/0002020].

\bibitem{Bonnet:2001uh}
F.~D.~Bonnet, P.~O.~Bowman, D.~B.~Leinweber, A.~G.~Williams and J.~M.~Zanotti,
Phys.\ Rev.\ D {\bf 64}, 034501 (2001)
[arXiv:hep-lat/0101013].

\bibitem{Langfeld:2001cz}
K.~Langfeld, H.~Reinhardt and J.~Gattnar,
Nucl. Phys. B {\bf 621}, 131 (2002)
[arXiv:hep-ph/0107141]; 
arXiv:hep-lat/0110025.

\bibitem{Suman:1995zg}
H.~Suman and K.~Schilling,
Phys.\ Lett.\ B {\bf 373} (1996) 314
[arXiv:hep-lat/9512003].

\bibitem{Cucchieri:1997dx}
A.~Cucchieri,
Nucl.\ Phys.\ B {\bf 508} (1997) 353
[arXiv:hep-lat/9705005].

\bibitem{Fischer:2003rp}
C.~S.~Fischer and R.~Alkofer,
Phys.\ Rev.\ D {\bf 67} (2003) 094020 [arXiv:hep-ph/0301094].

\bibitem{Kugo:1979gm}
T.~Kugo and I.~Ojima,
Prog.\ Theor.\ Phys.\ Suppl.\ {\bf 66}, 1 (1979).

\bibitem{Nakanishi}
N.~Nakanishi and I.~Ojima, 
{\it Covariant Operator Formalism of Gauge Theories and Quantum
Gravity}, World Scientific, 1990.


\bibitem{Baulieu:1982sb}
L.~Baulieu and J.~Thierry-Mieg,
Nucl.\ Phys.\ B {\bf 197} (1982) 477;
J.~Thierry-Mieg,
Nucl.\ Phys.\ B {\bf 261} (1985) 55.

\bibitem{Zwanziger:2002ia}
D.~Zwanziger,
arXiv:hep-th/0206053.

\bibitem{Szczepaniak:2001rg}
A.~P.~Szczepaniak and E.~S.~Swanson,
Phys.\ Rev.\ D {\bf 65}, 025012 (2002)
[arXiv:hep-ph/0107078];
\newline
C.~Feuchter, K.~Langfeld, L.~Moyaerts and H.~Reinhardt, to be published.

\bibitem{Cucchieri:2001zb}
A.~Cucchieri and D.~Zwanziger,
Nucl.\ Phys.\ Proc.\ Suppl.\  {\bf 106}, 694 (2002)
[arXiv:hep-lat/0110189];

A.~Cucchieri, T.~Mendes and D.~Zwanziger,
Nucl.\ Phys.\ Proc.\ Suppl.\  {\bf 106}, 697 (2002)
[arXiv:hep-lat/0110188];

A.~Cucchieri and D.~Zwanziger,
Phys.\ Lett.\ B {\bf 524}, 123 (2002)
[arXiv:hep-lat/0012024];

A.~Cucchieri and D.~Zwanziger,
Phys.\ Rev.\ D {\bf 65}, 014001 (2002)
[arXiv:hep-lat/0008026];

\bibitem{Cucchieri:2000hv}
A.~Cucchieri and D.~Zwanziger,
Phys.\ Rev.\ D {\bf 65}, 014002 (2002)
[arXiv:hep-th/0008248].

\bibitem{Baulieu:1998kx}
L.~Baulieu and D.~Zwanziger,
Nucl.\ Phys.\ B {\bf 548}, 527 (1999)
[arXiv:hep-th/9807024].

\bibitem{Zwanziger:2002sh}
D.~Zwanziger,
Phys.\ Rev.\ Lett.\  {\bf 90}, 102001 (2003)
[arXiv:hep-lat/0209105].


\bibitem{Greensite:2003xf}
J.~Greensite and S.~Olejnik,
arXiv:hep-lat/0302018.

\bibitem{Szczepaniak:1996tk}
A.~P.~Szczepaniak and E.~S.~Swanson,
Phys.\ Rev.\ D {\bf 55}, 3987 (1997)
[arXiv:hep-ph/9611310];

A.~P.~Szczepaniak and P.~Krupinski,
arXiv:hep-ph/0204249.

\bibitem{Alkofer:tc}
R.~Alkofer and P.~A.~Amundsen,
Nucl.\ Phys.\ B {\bf 306}, 305 (1988);


K.~Langfeld, R.~Alkofer and P.~A.~Amundsen,
Z.\ Phys.\ C {\bf 42}, 159 (1989);


R.~Alkofer and P.~A.~Amundsen,
Phys.\ Lett.\ B {\bf 187}, 395 (1987);

\bibitem{Lavelle:eg}
M.~J.~Lavelle and M.~Schaden,
Phys.\ Lett.\ B {\bf 208}, 297 (1988).

\bibitem{Boucaud:2001st}
P.~Boucaud, A.~Le Yaouanc, J.~P.~Leroy, J.~Micheli, O.~Pene and J.~Rodriguez-Quintero,
Phys.\ Rev.\ D {\bf 63}, 114003 (2001)
[arXiv:hep-ph/0101302].

\bibitem{Kondo:2001nq}
K.~I.~Kondo,
Phys.\ Lett.\ B {\bf 514}, 335 (2001)
[arXiv:hep-th/0105299].

\bibitem{Stodolsky:2002st}
L.~Stodolsky, P.~van Baal and V.~I.~Zakharov,
Phys.\ Lett.\ B {\bf 552}, 214 (2003)
[arXiv:hep-th/0210204].

\bibitem{Dudal:2003gu}
D.~Dudal, H.~Verschelde, V.~E.~Lemes, M.~S.~Sarandy, S.~P.~Sorella and M.~Picariello,
arXiv:hep-th/0302168.

\bibitem{Schaden:1999ew}
M.~Schaden,
arXiv:hep-th/9909011.

\bibitem{Gubarev:2000eu}
F.~V.~Gubarev, L.~Stodolsky and V.~I.~Zakharov,
Phys.\ Rev.\ Lett.\  {\bf 86}, 2220 (2001)
[arXiv:hep-ph/0010057].

\bibitem{Dan} D.\ Zwanziger, private communication.

\bibitem{Pierre} P.\ van Baal, private communication.

\bibitem{Gripaios:2003xq}
B.~M.~Gripaios,
arXiv:hep-th/0302015.

\bibitem{deBoer:1995dh}
J.~de Boer, K.~Skenderis, P.~van Nieuwenhuizen and A.~Waldron,
Phys.\ Lett.\ B {\bf 367}, 175 (1996)
[arXiv:hep-th/9510167].


\bibitem{Browne:2002wd}
R.~E.~Browne and J.~A.~Gracey,
Phys.\ Lett.\ B {\bf 540}, 68 (2002)
[arXiv:hep-th/0206111];
J.~A.~Gracey,
Phys.\ Lett.\ B {\bf 552}, 101 (2003)
[arXiv:hep-th/0211144].

\bibitem{Form}
J.~A.~Vermaseren,
arXiv:math-ph/0010025.

\bibitem{Watson:2001yv}
P.~Watson and R.~Alkofer,
Phys.\ Rev.\ Lett.\  {\bf 86}, 5239 (2001)
[arXiv:hep-ph/0102332];
R.~Alkofer, L.~von Smekal and P.~Watson,
Proceedings of the ECT* Collaboration Meeting on Dynamical Aspects of the QCD 
Phase Transition, Trento, Italy, March 12-15, 2001, 
arXiv:hep-ph/0105142.

\end{thebibliography}
\end{document}